\documentclass{article}
\usepackage{arxiv}
\usepackage[utf8]{inputenc}
\usepackage{arabtex}
\usepackage{graphicx}
\usepackage{adjustbox}
\usepackage{utf8}
\usepackage{multirow}
\usepackage{hyperref}
\usepackage{makecell}
\usepackage[table]{xcolor}
\usepackage{subcaption}
\usepackage{amsmath}
\usepackage{cleveref}
\usepackage{tabularx}
\usepackage{xcolor}
\usepackage{color, soul}
\title{Eating Garlic Prevents COVID-19 Infection: Detecting Misinformation on the Arabic Content of Twitter}
\author{
 Sarah Alqurashi\thanks{Corresponding author: \texttt{s43980127@st.uqu.edu.sa}} \\
  Center of Innovation and Development in Artif. Intell. (CIADA)\\
  Umm Al-Qura University\\
  Makkah, Saudi Arabia \\
  \AND
  Btool Hamoui \\
 Center of Innovation and Development in Artif. Intell. (CIADA)\\
 Umm Al-Qura University\\
 Makkah, Saudi Arabia \\
  \AND
 Abdulaziz Alashaikh \\
  Computer and Networks Engineering Department\\
  University of Jeddah\\
  Jeddah, Saudi Arabia \\
  \AND
  Ahmad Alhindi \\
  Center of Innovation and Development in Artif. Intell. (CIADA) \\
  Umm Al-Qura University\\
  Makkah, Saudi Arabia \\
    \AND
  Eisa Alanazi \\
  Center of Innovation and Development in Artif. Intell. (CIADA) \\
  Umm Al-Qura University\\
  Makkah, Saudi Arabia \\
}

\begin{document}

\maketitle
\begin{abstract} 
The rapid growth of social media content during the current pandemic provides useful tools for disseminating information which has also become a root for misinformation. Therefore, there is an urgent need for fact-checking and effective techniques for detecting misinformation in social media. In this work, we study the misinformation in the Arabic content of Twitter. We construct a large Arabic dataset related to COVID-19 misinformation and gold-annotate the tweets into two categories: {misinformation or not}. Then, we apply eight different traditional and deep machine learning models, with different features including  word embeddings and word frequency. The word embedding models (\textsc{FastText} and word2vec) exploit more than two million Arabic tweets related to COVID-19. Experiments show that optimizing the area under the curve (AUC) improves the models' performance and the Extreme Gradient Boosting (XGBoost) presents the highest accuracy in detecting COVID-19 misinformation online.
 \end{quote}
\end{abstract}
\section{Introduction}
The new coronavirus pandemic was accompanied by a large and rapid spread of rumors, false information, and fake news. Misinformation has existed over the years and usually flourish on various important issues such as health outbreaks, climate change, and vaccinations. Human crises are fertile ground for misinformation, as it happened during the Zika virus \cite{ghenai2017catching}, Ebola \cite{oyeyemi2014ebola}, and others. Moreover, misinformation is intensified during sudden and intense crises such as the COVID-19 pandemic. However, in the modern era, social media has helped magnify the spread of misinformation among individuals. As in recent times, there has been a global increase in the spread of information in general, especially misinformation related to COVID-19 through various social media. The unprecedented amount of information poses serious public health challenges, especially concerning infectious diseases, which prompted the World Health Organization (WHO) to warn against the infodemic. The infodemic is a massive amount of correct and incorrect information, making it difficult for individuals to access reliable information and credible guidance when needed \cite{infodemic}. This phenomenon, in turn, leads to a fast and easy spread of fake and unreliable information, especially on social media, which facilitates the diffusion of misinformation.

Several conspiracy theories about the origins of the COVID-19 virus have spread on Arabic social media, all with a common idea that the virus was a biological weapon. This misinformation started from social media accounts with no reliable proof to back their claims. Moreover, misleading information about the virus's symptoms and how to cure the new virus and reduce its transmission circulates on social media. For example, a widespread misinformation claimed that home remedies such as taking vitamin C and eating garlic can treat and prevent COVID-19 infection with a complete lack of evidence. Although some home remedies are harmless, some can be very dangerous. While these misinformation serve their promoters' interests, it also harms societies. Especially that a high percentage of individuals depend on a social media platform for information and news. Research has shown that the more individuals are exposed to false information and fake news, the more likely they will accept and believe it \cite{del2016spreading}.  Misinformation confuses people and causes harm to the health of individuals. It may also incite violence, discrimination, or hostility against specific groups in society. Furthermore, it may obstruct the efforts to control the current health crisis.

Twitter is one of the most used social networking sites in the Arab world that has become a tool for spreading misinformation regarding COVID-19.  A recent study shows that false information spreads six times faster than correct information on Twitter \cite{MIT}, which makes it challenging to find accurate information on Twitter, causing an increase in mental distress and anxiety during the pandemic. One of the misinformation concerning factors is the spread  rate  on Twitter exceeds physical distances. The early spread of conspiracy theories and other false and misleading information may occur on Twitter. However, it may reach a larger audience once it appears, and it may be amplified by social media influencers as well through reports in unreliable media sites, which reduces the effectiveness of officials attempts in slowing the spread of such misleading information. As a result, individuals around the world are affected mentally and physically by misinformation.  
The World Health Organization has teamed up with prominent social media platforms such as Facebook, Twitter, and YouTube to fight the infodemic by verifying irrelevant information and providing evidence-based information to the public \cite{donovan2020here}. Despite the efforts made by different entities around the globe, including WHO, governments, and social media sites, misinformation continues to spread widely. The problem lies in the difficulty of detecting and correcting misinformation in the Arabic content before it spreads more widely. The Arabic language also poses a challenge because it has many dialects and rich vocabulary, which makes misleading information exists in more than one dialect making it harder to detect. As a result, there is an urgent need to develop systems that are capable of automatically identifying misinformation in Arabic content. 
In this work, we investigate detecting Arabic misinformation on Twitter using natural language processing and machine learning. Our contributions to this area are summarized as follows:
\begin{itemize}
\item We extract a sample of tweets from a large Arabic dataset related to the COVID-19 pandemic. Human annotators are utilized for labeling the sample. With high-quality, human-powered data annotation, we can estimate the credibility of the considered tweets automatically. 
\item We build two Arabic word embedding models using \textsc{FastText} and word2vec based on more than two million Arabic tweets related to COVID-19 for a comparative analysis between the classifiers. 
\item We examine the prediction performance on five traditional classifiers: Random Forests (RF), Extreme Gradient Boosting (XGB), Naive Bayes (NB), Stochastic Gradient Descent (SGD), and Support Vector Machines (SVM) with different features in addition to three other deep learning classifiers CNN, RNN, and CRNN.
\item We improve the performance of all the models by optimizing the area under the curve using grid search for the traditional classifiers and AUC loss function for the deep learning models.
\end{itemize}

\section{Misinformation Analysis}
To reduce misinformation on social media, it is essential to understand what the term misinformation means. Some scholars describe \emph {misinformation} as false and inaccurate information that unintentionally transmits \cite{persily2020social}. Usually, ordinary users spread this type of misinformation because of their confidence in the information source, whether they were personally acquainted with or were influential users on their social network. They share the misinformation to inform people in their surroundings about a specific situation or story because they believe it is true. 

In contrast, \emph{disinformation} is known as false and inaccurate information that is transmitted intentionally \cite{persily2020social}. 
Usually carried by a group of people/writers or even publishers with a common goal to deceive the public and promote disinformation. 
Disinformation includes conspiracy theories, fake news, and spams. The outcome of mis- and dis-information is the same, whether it is published intentionally or not.

On social media where users can post anything, it is difficult for researchers to determine whether a piece of information was intentionally created or not. Therefore, misinformation has been identified as an umbrella term for all false and inaccurate information, regardless of the goal or intention \cite{wu2019misinformation}.
The umbrella term misinformation includes fake news, which is a type of misinformation that mimics traditional news, rumors, which are unverified information that can be correct, and spams, unwanted information that exhaust its recipient \cite{wu2019misinformation, islam2020deep}.  These misinformation types share a negative impact, as their impact extends on every aspect of life, and may have social and economic consequences. Furthermore, misinformation has a significant impact on emergency response during disasters. It aims to mislead and confuse the public opinion and threaten public security and community stability, especially in the absence of immediate intervention to combat it \cite{islam2020deep}.


\section{Related Work}
Due to social media ease of use, the spread of misinformation has expanded across ranges. The impact of misinformation goes beyond personal life to affecting society and even the economy. One of the examples of misinformation that has a negative effect is the spread of inaccurate information related to vaccinations. Anti-vaccinations groups claim that vaccines cause autism, which caused fear of vaccinations among many parents, making them refuse or at least hesitate to vaccinate their children, which caused an unprecedented increase in preventable diseases \cite{lewandowsky2012misinformation}. The fear of vaccination continue during the global pandemic of COVID-19, as some conspiracy theories spread through social media platforms claiming that the COVID-19 vaccine contained a chip that controls humans. 

The amount of data on social media makes it difficult to distinguish between misleading and accurate information. Therefore, identifying misinformation via social media has been a popular topic in recent years. 

{ Many studies in the English language have examined the presence of misinformation on social media, such as detecting rumors \cite{akhtar2018no}, fake news \cite{buntain2017automatically}, spam \cite{wang2010don} and heath misinformation \cite{ghenai2017catching,oyeyemi2014ebola}. However, most of the Arabic language research focused on identifying information credibility of news disseminated in Twitter. Often, the tweets were annotated based on an annotator judgment and machine learning models are used based on user or content features, or a combination of both features \cite{al2010measuring,hassan2018supervised}. Some studies added new features such as sentiment analysis \cite{jardaneh2019classifying,el2017cat}, user replies polarity\cite{sabbeh2018arabic}, the similarity between username and display name \cite{mouty2019effect}, and TF-IDF \cite{hassan2020credibility}. The work in \cite{alzanin2019rumor} used content and user features to detect Arabic rumor from Twitter using semi-supervised expectation-maximization (E-M). The proposed model achieved an F1 score of 80\%.  However, little work so far has focused on detecting and tracking health misinformation in the Arabic language. Recently, a study that tackled the problem of detecting Arabic cancer treatment-related rumors on Twitter was presented in \cite{saeeddetecting}. They utilized ten machine learning models using TF-IDF features with different n-grams extracted form a dataset of 208 annotated tweets. An oversampling technique was applied to the dataset where it achieved F1 score of 0.86 by the random-forest model with oversampling and 5 gram TF-IDF features.}

There is a great body of work related to COVID-19 infodemic in social media. The evolution of misinformation was studied on the Weibo social media platform \cite{leng2020analysis} using misinformation identified by fact-checking platforms. Another study \cite{serrano2020nlp} examines the identification of misinformation videos on YouTube using NLP and machine learning. Furthermore, the work in \cite{ng2020pofma} presented an analysis of the evolution of the opinion of Singapore telegram group chat regarding COVID-19.

The vast majority of COVID 19 infodemic studies on social media platforms focused on Twitter largely because Twitter is one of the most popular social media platforms. Twitter also provides access to a large amount of content in many languages. Along this line, many studies of misinformation on Twitter focused on analyzing the content of tweets to understand Twitter conversion during COVID 19 \cite{singh2020first,mourad2020critical,medford2020infodemic}. To study the development of conversation around misinformation on Twitter, Singh et al.\cite{singh2020first},  collected five common misinformation related to COVID-19, which are about the virus's origin, vaccine development, flu comparison, heat kills the disease, and home remedies. Each tweet is assigned to corresponding misinformation based on words and phrases in the tweets. The authors noticed an increase in the conversation around the misinformation since January 2020. In  \cite{pulido2020covid}, a study of disseminating COVID 19 misleading information and reliable information on Twitter using communicative content analysis shows that the likelihood of misleading information to be retweeted is less than accurate information.

Several studies have relied on fact-checking websites as ground truth data. In \cite{shahi2020exploratory}, authors collected the COVID-19 related tweets that have been mentioned in fact-checking articles to study the source of misinformation and how it is spreading. The retweet speeds were used as a proxy for the propagation speed of misinformation. Their work suggests that the propagation speed of misinformation is higher than accurate information. Another study on how misinformation content spreads over five months on Twitter was presented in \cite{mcquillan2020cultural}. 
On a different note, the work in \cite{mourad2020critical} presents a  different measure of the tweets' credibility based on user specialty and occupation.  

Considerable work has also focused on the quality of the links and information sources found in tweets in many languages (e.g., English \cite{singh2020first}, Italy \cite{caldarelli2020analysis}). The links were examined and classified as reputable sources or not, using fact-checking websites \cite{caldarelli2020analysis} and well-known domains \cite{singh2020first}. Low-quality links were less used in tweets than high-quality links.

Researchers also studied the type of accounts that help spread false information about COVID-19. The role and behavior of bot accounts on Twitter during COVID-19 were analyzed in \cite{graham2020like,ferrara2020types} where it was shown that Twitter bots participate in misinformation propagation on Twitter, either for political or marketing gain \cite{graham2020like}.

Machine learning techniques have also been adapted to detect misinformation. The work in \cite{ding2020challenges} discussed the challenges in designing and developing an AI solutions for infodemic detection. Moreover, authors presented a tool to estimate whether an  article is a misinformation based on URL checker, fake news classifier, and website matcher.
A misleading information detection system was presented in \cite{elhadad2020detecting}. The system relies on the fact-checking website and international organization data. The system is based on ensemble techniques built using 10 machine learning models with 7 feature extraction techniques. Another study in \cite{al2020lies} deploys machine learning techniques on Twitter misinformation using ensemble techniques based on user level and tweet level features. The models that had great accuracy were SVM and random forest.

Most of the research has focused on the English language. However, there are very few studies on the Arabic language. In \cite{alam2020fighting}, they applied SVM, \textsc{FastText}, and BERT on 218 Arabic tweets and 504 English tweets. The \textsc{FastText} model provided the best result for Arabic text. The work in \cite{alsudias2020covid} study the Arabic conversation on Twitter by applying topic modeling. Machine learning models such as logistic regression, support vector machines, and naive Bayes were used on 2000 labeled tweets to build rumors detection system. The highest accuracy is 84\% achieved by logistic regression classifier with vector count features. Also, they found out that rumors are usually written in an academic way and promoted by fake health professionals. Another study of COVID-19 misinformation was presented in \cite{mubarak2020arcorona}. The study published a large, manually annotated dataset of Arabic tweets related to COVID-19. The tweets were labeled based on 13 classes, including only 421 rumors. The author employed machine learning and transformer models using Mazajak embeddings and TF-IDF n-gram for words and characters. The best model was SVC with TF-IDF characters n-gram with  0.79 F1 scores. 
\begin{table}[ht]
\begin{center}
\renewcommand{\arraystretch}{2}
\begin{adjustbox}{width=\textwidth}
\begin{tabular}{ |c|c|c|c|c| }
\hline
 Paper & Purpose & \#Arabic Tweets & Features & Classifiers \\ 
 \hline
\cite{hassan2020credibility} & Assessing Arabic and English tweets credibility  & 9000 & TF and TF-IDF & Traditional classifiers  \\  
\hline
\cite{sabbeh2018arabic} & Assessing Arabic news credibility in Twitter  & 800 &\makecell{User, content, and replies polarity \\ based features } & Traditional classifiers  \\  
\hline
\cite{al2010measuring} & Assessing Arabic tweets credibility  & 600 &\makecell{user, content \\ based features} & Traditional classifiers  \\  
\hline
\cite{hassan2018supervised} & Assessing Arabic tweets credibility  & 5802 & \makecell{user, content \\based features} & Traditional classifiers  \\  
\hline
\cite{jardaneh2019classifying} & Assessing Arabic news credibility in Twitter  & 1862 &\makecell{user, content and \\ sentiment based features} & Traditional classifiers  \\ 
\hline
\cite{el2017cat} & Assessing Arabic tweets credibility & 9000 & \makecell{User, content and \\ sentiment based features} & Traditional classifiers  \\ 
\hline
\cite{mouty2019effect} & Assessing Arabic tweets credibility & 9000 & \makecell{user, content, and  similarity between \\ username   and display name based features }& Traditional classifiers  \\ 
\hline
\cite{alzanin2019rumor} & Detecting  Arabic rumor on Twitter   & 177 & User and content based features & Semi-supervised expectation-maximization (E-M)  \\  
\hline
\cite{saeeddetecting} & Detecting cancer treatment rumors & 208 & TF-IDF n-gram & Traditional classifiers  \\  
\hline
\cite{alam2020fighting} & Detect COVID19 misinformation & 218 & TF-IDF n-gram & SVC,\textsc{FastText} and BERT  \\  
\hline
\cite{alsudias2020covid} & Detect COVID19 misinformation & 2000 &\makecell{ Count vector ,TF-IDF, and \\ word embedding }& Traditional classifiers \\   
 \hline
\cite{mubarak2020arcorona} & Analyzing COVID19 Arabic Tweets & 8000 & \makecell{TF-IDF n-gram for words and \\ characters} & Traditional classifiers \\   
 \hline
Our work  & Detect COVID19 misinformation & 8786 & \makecell{TF-IDF, FastText,\\ and word2vec word embedding} &  Traditional  and deep learning classifiers\\ 
 \hline
\end{tabular}
\end{adjustbox}
\caption{Summary for different attempts in detecting misinformation from the Arabic content of Twitter.}
\label{tb10:covid}
\end{center}
\end{table}

In all previous studies investigating the Arabic content of misinformation on social media, the used datasets were very limited. In this work,  we construct one of the largest datasets of tweets for misinformation in Arabic language. We provide a comparative analysis between the classifiers based on using TF-IDF and Arabic word embedding models, built based on more than two million Arabic tweets related to COVID-19. Furthermore, we further optimize the area under the curve (AUC) to better improve the models accuracy. 
%
\section{Methodology}
\begin{figure}[ht] 
    \centering
    \includegraphics[width=.9\linewidth]{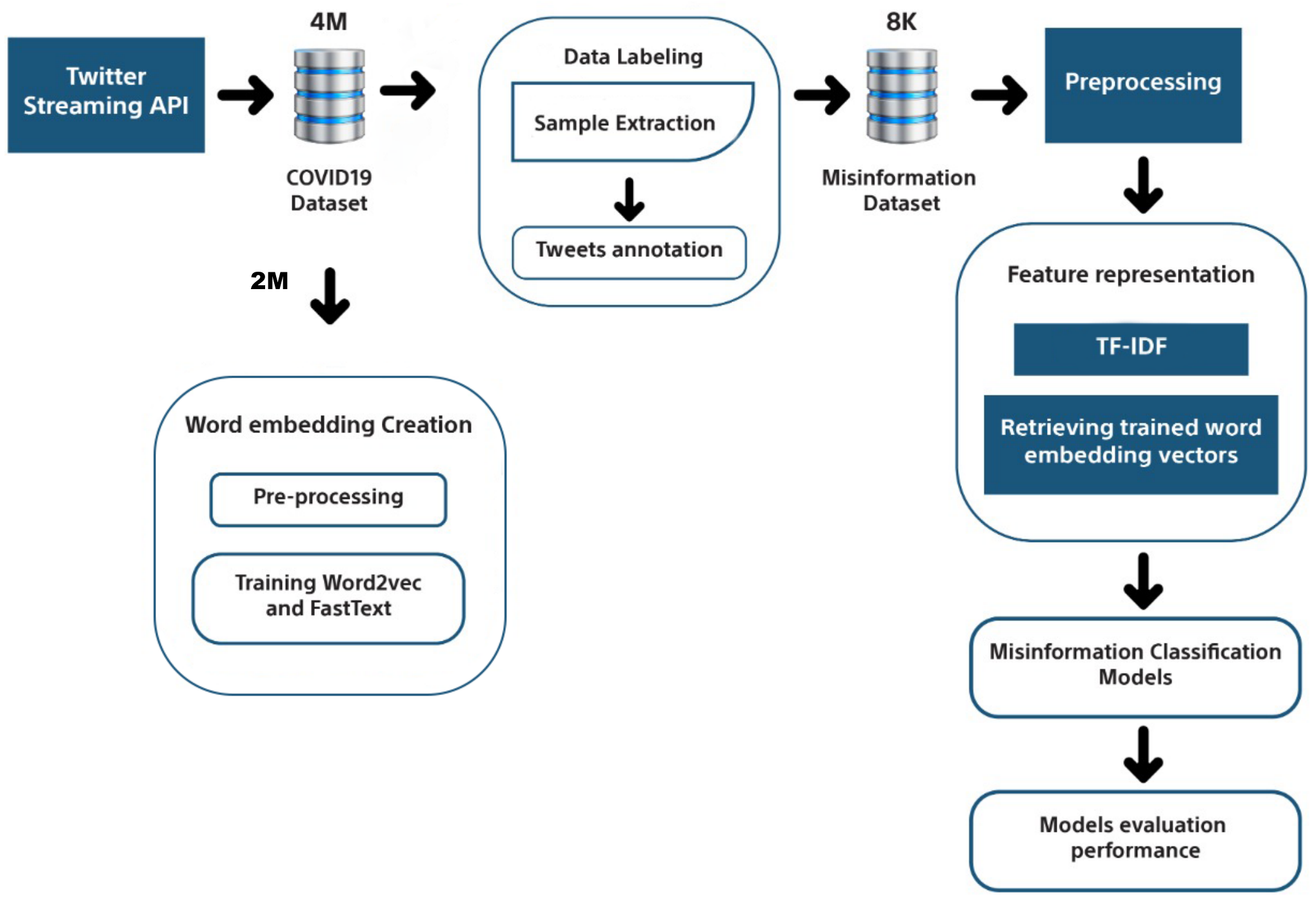}
    \caption{System architecture.}
    \label{fig:meth}
\end{figure}

The proposed system comprises several stages shown in \Cref{fig:meth}. It begins by collecting tweets using the Twitter streaming API and ends with evaluating the models performance and comparing them. 
In the remaining of this section, we describe the steps in details. 
\subsection{Data Collection}
We collected a large number of Arabic tweets using the Twitter streaming application interface and Tweepy Python library for four months from January 1, 2020, to April 30, 2020. We extracted tweets based on a list of the most common Arabic keywords associated with COVID-19. We filter the Twitter stream based on the Arabic language and obtain relevant tweets about the pandemic. \Cref{tb9:keywords} shows a list of all the relevant Arabic keywords used to collect tweets about COVID-19. The dataset contains more than 4,514,136 million tweets. We store the tweet's full object, including the timestamp of the tweet, the id of the tweet, user profile information including the number of followers, and geolocation of the tweet in a MongoDB NoSQL database. The dataset is available online on GitHub \footnote{\url{https://github.com/SarahAlqurashi/COVID-19-Arabic-Tweets-Dataset}}. 

\begin{table}[htb]
\begin{center}
\footnotesize
\begin{tabular}{|c|c|c|}
\hline
 \textbf{\textit{Keyword}}& \textbf{\textit{English Translation}} &\textbf{\textit{Tracing Date} } \\
\hline
\setcode{utf8}
\<الفايروس التاجي>
& Coronavirus & 2020-01-01 \\
\hline
\<كورونا> 
&Corona & 2020-01-01\\
\hline
\<ووهان> 
 & Wuhan& 2020-01-01\\
\hline
\<الصين> 
 & China& 2020-01-01 \\
\hline
\<تفشي> 
 & Outbreak & 2020-01-01\\
 \hline
\<كمامة >
 & Mask & 2020-01-01\\
\hline
\<كمامات> 
 & Masks & 2020-01-01\\
 \hline
 \<معقمات> 
 & Sterilizers& 2020-01-01 \\
 \hline
\<تعقيم> 
 & Sterilization & 2020-01-01\\
\hline
\<غسل اليدين> 
 & Washing hands & 2020-01-01\\
\hline
\<العزل المنزلي>
 & Home isolation & 2020-02-01\\
\hline
\<جائحة> 
 & Pandemic & 2020-01-22\\
 \hline
\<وباء> 
 & Epidemic & 2020-01-22\\
\hline
\<الحجر المنزلي>
& Home quarantine & 2020-02-01\\
\hline
\<كوفيد 19> 
 & COVID 19& 2020-03-01\\
 \hline
\<حظر التجول> 
  & Curfew & 2020-03-15\\
 \hline
\<التباعد الاجتماعي> 
  & Social distancing  & 2020-04-01\\
 \hline
\< جهاز التنفس الصناعي> 
  &  Ventilator & 2020-04-01\\
 \hline
\< ضيق تنفس> 
  & Shortness of breath  & 2020-04-01\\ 
\hline
\< كحة> 
  & Cough  & 2020-04-01\\ 
\hline
\< حراره> 
  & temperature  & 2020-04-01\\
\hline
\< متر ونص> 
  & One and a half meters & 2020-04-01\\ 
\hline
\< فعاليات الحجر> 
  & Quarantine activities  & 2020-04-01\\ 
\hline
\< الحجر الصحي> 
  & Quarantine  & 2020-04-01\\ 
  \hline
\< دواء الملاريا> 
  & Malaria medicine  & 2020-04-25\\ 
  \hline
\< رمديسيفير> 
  & Remdesivir  & 2020-04-25\\ 
  \hline
\< رفع الحظر> 
  & Curfew lift  & 2020-04-26\\
  \hline
\< الحظر الجزئي> 
  & Partial curfew  & 2020-04-26\\
  \hline
\<المسح المجتمعي> 
  & Active surveillance  & 2020-04-29\\
  \hline
\< الفحص النشط> 
  & Active testing  & 2020-04-29\\
\hline
\end{tabular}
\end{center}
\caption{The list of keywords that we used to collect the tweets.}
\label{tb9:keywords}
\end{table}
\subsection{Data Prepossessing}
When dealing with Arabic data, it is important to recognize the rich cultural and linguistic diversity across the Arab region which may translate into different challenges (e.g., dialects) that must be addressed during the model's development. It is also essential to consider the general features of the Twitter data. For example, tweets size is limited to 280 characters. Despite this, the content of the tweets are varied and can consist of texts, symbols, URLs, pictures, and videos. Furthermore, on Twitter, users tend to use informal writing methods to reduce the text's length, while others can still comprehend it. Also, Twitter data contains large amounts of spelling errors and does not necessarily follow the language's formal structure. Thus, Twitter data becomes very noisy. Accordingly, it is essential to apply some pre-processing to the raw text data before feeding it to the classifiers. We perform the following preprocessing steps to the tweets: 
\setcode{utf8}
\begin{itemize}
\item We removed non-Arabic words.
\item We removed special characters such as (\#, \%, \&, \@).
\item We removed URLs.
\item We removed Arabic diacritics and punctuation marks (\< ُ , . ,! ,?>).
\item We performed text correction using Textblob python Library \cite{TextBlob}
\item We Normalized Arabic text by :
Replacing (\<أ , إ , آ  , ا >)  with (\<ا> ) and
Replacing (\<ئ >)  with (\<ا> )   and 
Replacing (\<ى >)  with (\<ي> ) and 
Replacing (\<ة >)  with (\<ه> ) and 
Replacing (\<ؤ >)  with (\<و> ) and 
Replacing (\<ة >)  with (\<ه> ) and 
Replacing (\<گ >)  with (\<ك> )  
\item We removed  the repetition of characters such as (\<عاااااجل  >) turns into  (\<عاجل  > )
\item We removed stop words such as  (\<على ، إلى ، في   > 
)  (from, to , in ).
\item  We performed word stemming to convert each word to  its corresponding root  using Farasapy library \cite{farasapy}
\end{itemize}
{\subsection{Data Labeling}
\textbf{Misinformation identification:}  This work copes with health-related misinformation detection. We do this by relying on trusted sources of information. A recent work \cite{alqurashi2020identifying} shows that the official account of the Ministry of Health in Saudi Arabia was among the top influential accounts in March 2020. Hence, we collected the false information reported on both the World Health Organization (WHO) website and the Ministry of Health in Saudi Arabia website. \Cref{tb10:Examples} shows a sample of tweets containing misinformation.

\textbf{Dataset annotation:} our misinformation dataset is sampled from tweets collected from early March 2020 to the end of April 2020. To narrow down the set of tweets without misinformation content, we use the similar procedure as used by \cite{ghenai2017catching}. We first manually crafted a set of terms that best describe different misinformation. Then, we retrieved tweets related to those terms 
(e.g. “Vitamin C: \<فيتامين سي>”, “Sarin gas:\<غاز السارين>”, “Mosquitoes: \<البعوض>”, and “Biological warfare: \<حرب بيولوجية>”).  The tweets were then combined into one dataset and labelled by two Arabic native speaker volunteers. Before labeling the tweets, the annotators reviewed the list of the collected misinformation. Due to the substantial manual effort involved in labeling these tweets, each tweet in the dataset was labeled by exactly one annotator. The tweet which contains misinformation were labeled  "1" and others were labeled by "0".      

In total, our misinformation dataset consists of 8,786 Arabic tweets, which contains 36,198 unique words after applying pre-processing. Overall, our labelled misinformation dataset covers significant misleading and inaccurate content that were widely circulated among Arabic tweeters during March and April. The number of tweets containing misinformation in April (709 tweets) was higher than its counterpart in March (602 tweets). \Cref{tb5:Dataset} shows general statistics about the dataset. Recall that we consider a (Misinformation) as the tweet that has been labeled by “1” and  (Other) as the tweet that has been labeled by “0” by the annotators. From \Cref{tb5:Dataset}, we observe that the dataset is unbalanced; the majority class (Other) has 7,475 more tweets than the minority class (Misinformation). The misinformation data-set is freely accessible on GitHub. \footnote{\url{https://github.com/SarahAlqurashi/COVID19-Misinformation-dataset-}}.
\begin{table}[ht]
\begin{center}
\begin{adjustbox}{width=\textwidth}
\begin{tabular}{|l|ll|}
\hline

\textbf{Misinformation headline} & \multicolumn{2}{l}{\textbf{Tweet examples}} \\
\hline
\multirow{2}{*}{\makecell{Eating garlic protects \\ against coronavirus}}& Ar. &
\< يساعد تناول الثوم  على منع الاصابه بفيروس كورونا لتقويه المناعه  في جسم الانسان >\\           
& En. & Eating garlic helps prevent coronavirus infection, and to   strengthen the body’s immunity. \\

\multirow{2}{*}{\makecell{ Gargling with salt protects\\ against corona virus}} & Ar. &                  \< التعرض للشمس والغرغره بالماء الدافئ والملح طرق بسيطه لحمايتك كورونا >                \\
& En. & Exposure to the sun and gargling with warm water and salt   are simple ways to protect you from coronavirus   \\

\multirow{2}{*}{\makecell{Coronavirus is sarin\\  gas}} & Ar. & 
\< المنتشر بالجو غاز السارين وليس كورونا مده انتشاره بالهواء اشهر وموعد انقضاء فتره بقااه بالجو ستنتهي تاريخ >  \\
& En  & Sarin gas are spread in the atmosphere and not  coronavirus, its diffusion in the air is months,\\

\multirow{2}{*}{\makecell{Mosquitoes transmit\\ infection}}   & Ar. &  
\< مصادر متينه يثبت نقل الحيوانات الاليفه والبعوض لفيروس كورونا > \\        
& En  & trustable sources transferee pets and mosquitoes to the   coronavirus     \\ 

\multirow{4}{*}{\makecell{5G Networks Spreading \\ the Coronavirus}}& Ar. &
\< في ووهان الصينيه شبكة  5 ساعدت في نشر الفيروس بصوره كارثيه مما تسبب في نقص الاوكسجين في اجسامهم  >   \\          

 & &
\<في الشوارع ادركت الصين ذلك فرجعوا لاستخدام شبكة 3 في هواتفهم>\\

 & En  & \multirow{2}{*}{\makecell{ In China Wuhan, Network 5G helped spread the virus in a catastrophic manner, causing a lack of oxygen in Human\\ bodies and falling onto the streets. China realized this, so they returned to using Network 3G in their phones } } \\
&  & \\

\hline
\end{tabular}
\end{adjustbox}
\end{center}
\caption{Examples of Misinformation in tweets}
\label{tb10:Examples}
\end{table}
\begin{table}[ht!]
\begin{center}

\begin{tabular}{|l|l|l|l|l|}
 \hline
        & Misinformation & Other & Total & \#words  \\
 \hline        
Dataset &  1,311     & 7,475 & 8,786 & 36,198 \\  
 \hline
\end{tabular}
\end{center}
\caption{Statistics of the dataset.}
\label{tb5:Dataset}

\end{table}

\subsection{Feature Representation}
This step involves transforming the pre-processed tweets texts into the feature vector where we construct the feature vector for each tweet from a vectorization or word embedding. 
In that sense, from tokenized vectors of words, we build the feature vectors using TF-IDF \cite{salton1988term} and word embeddings techniques \cite{mikolov2013efficient,bojanowski2017enriching}. 
The concept behind each of them is briefly explained as follows:
\begin{itemize}
\item\textbf{Sparse Vector Based on TF-IDF:}
In this representation, the importance of a term/(n-gram) in a tweet is evaluated in relation to the whole dataset. This representation method gives high weights to terms that are specific to some tweets and decreases the weight of frequently occurring words in the whole dataset. It is composed of term frequency (TF) and inverse document frequency (IDF) and it is computed using Equation \ref{TFIDF}: 
\begin{equation}
\label{TFIDF}
TF-IDF(W_ij) = TF_i  x \log (\frac{N}{DF_i})
\end{equation}

where, \(TF_i\) is the number of occurrences of $i$ in $j$, $N$ is the total number of tweets, and \(DF_i\) is the number of tweets containing the word $i$. We constructed TF-IDF vectors twice; once with unigrams and another with N-grams. We obtained for each tweet a sparse vector of dimension 5000 with both unigrams and N-grams and used the Scikit-learn tool for the implementation.

\item\textbf{Word Embeddings Creation:}
In natural language processing, word embedding refer to the techniques used in mapping words or phrases to vectors of real numbers.
Word embedding methods represent words as continuous vectors in a low
dimensional space. These vectors capture semantic information between words; the words with similar meaning will have vectors closer to each other.

Building word embedding model using a large-scale training dataset is important to obtain meaningful embeddings \cite{elrazzaz2017methodical}. We built a word vectors model exploiting our whole COVID-19 dataset collected from January 2020 to April 2020. By removing retweets and duplicated tweets, we ended with 2,821,940 tweets. We consider two noticeable word embeddings generation methods: word2vec, and \textsc{FastText}. To train our models, we adopt the pre-processing pipeline of \cite{zahran2015word}. We investigate the following two types of word embeddings in this work:
\begin{itemize}
\item word2vec \cite{mikolov2013efficient}: This is probably the most widely used technique to learn word embeddings utilizing a shallow feed forward neural network. To build word2vec model, we take into consideration the maximum length of a tweet is 280 characters, hence, we use small context window sizes $W=3$. The model trained using the CBOW algorithm with dimension $D=200$. For the rest of parameters, we set the batch size to $50$, negative sampling to $10$, minimum word frequency to $5$ and iterations to $5$.  

\item \textsc{FastText} \cite{joulin2016FastText}: in \textsc{FastText}, the smallest unit is character-level n-grams and each word consists of a bag of character n-grams. This representation helps capture the meaning of shorter words and 
provides extraction of all prefixes and suffixes of a given word. For this reason, \textsc{FastText} has been shown to be more accurate and effective comparing to word2vec \cite{joulin2016bag}. To train \textsc{FastText} model, we used a small window of size $3$ with a dimension of $200$. We set minimum word frequency to $5$ and iterations to $5$. 

It is worth noting that very few studies have been developed \textsc{FastText} models for the Arabic language.The general word embeddings model such as the \textsc{FastText} model developed by \cite{bojanowski2017enriching} was trained on the Arabic Wikipedia Articles written in Modern Standard Arabic (MSA), hence employing such a model will not perform well on Twitter datasets. In the literature, the Arabic misinformation studies that employed \textsc{FastText} model, used \textsc{FastText} model in an unsupervised manner to produce feature vectors in \cite{alsudias2020covid}  while it was used in a supervised manner to predict the class labels in \cite{alam2020fighting}.However, there is a lack of detail about the training set sizes and parameters used in training the word embeddings model used in \cite{alsudias2020covid}.A recent study showed that unsupervised pre-training \textsc{FastText} on domain-specific can improve the classification quality over the supervised one, particularly when the dataset labels are limited\cite{agibetov2018fast}.
Hence, we opted to employ the \textsc{FastText} method where models are pre-trained through unsupervised training in our classification models.

\end{itemize}
We use the Gensim \cite{rehurek_lrec} implementation for the word2vec and \textsc{FastText} tools and we use the scheme that was used by \cite{yang2018using} to build the tweet-level representation for the machine learning models. Given the word2vec or \textsc{fasttex} models, we retrieve the vector representation of each word in each tweet, by averaging the word vectors of all words per tweets, as follow:
\begin{equation}
\label{ave_wordvectors}
V_{tweet2vec} =  \frac{\sum_{i=1}^{n} W_i}{n}
\end{equation}
where $n$ is the number of words in the tweet, and $W_i$ is the word2vec embedding for the $i$ word. This representation retains the number of dimensions ($D = 200$) in the word embedding models.
The word embeddings models used in our work is made freely available \footnote{available at: \url {https://github.com/BatoolHamawi/COVID-19WordEmbeddings}}.
\end{itemize}}
\subsection{Machine Learning Models}
To automatically predict COVID-19 misinformation in the Arabic Twitter, we used different types of classifiers. In this section, we will present them in detail.
  The first type of classifiers includes traditional (e.g., not deep) classifiers which are: support vector machine (SVM), multinomial naive Bayes (NB), Extreme Gradient Boosting (XGBoost), Random forest (RF), and Stochastic Gradient Descent (SGD). We used the implementation of these classifiers from the scikit-learn library \cite{varoquaux2015scikit}. 

The second type are deep learning models. The deep learning classifiers involved a convolutional neural network (CNN),  Recurrent Neural Networks with bidirectional long short-term memory (RNN BiLSTM), Convolutional Recurrent Neural Networks (CRNN). We used the implementations of these classifiers in Pytorch \cite{paszke2019pytorch}. Each proposed deep learning model consists of an input embedding layer, hidden layer, a dense output layer, and an  activation function. The embedding layer is the first layer of our deep learning models, and it creates a dense vector representation from the inputted text sequence. It can be initialized by a pre-train word embedding model or learned while training the model.  We experiment with three types of embedding layers to train the models. In the first experiment, the weights of the embedding layer are initialized randomly, and it will learn the embedding for all the words in the dataset. The second and third embedding layers are initialized using the weights from pre-trained word2vec and \textsc{FastText}.  Once the embedding layer maps each sequence text into a vector representation, the embedding representation is fed into the classifiers. The dense output layer takes the number of categories available as its output dimension.  We used the sigmoid activation function and cross-entropy loss function. In the following, we describe the models structures .
\begin{itemize}
 \item \textbf{ Convolutional Neural Network (CNN)}: In the CNN model, we use a one-dimensional convolution layer with a multi-scale kernel 4 and 5 with a fixed length of 100  for each filter dimensionality. The kernel size defines the number of words to consider as the convolution passes over the word vector resulting in different n-grams. The application of a convolution operation using one filter window over the word vector produces a new features map. After each convolution operation, we apply a nonlinear transformation using a Rectified Linear Unit (ReLU) \cite{nair2010rectified}. The convolved result is pooled using the maximum pooling operation to capture the text's most relevant features.  Then all feature maps are concatenated in one single vector with a fixed length.  Finally, we feed this vector through a fully-connected layer with a 0.5 dropout rate.

 \item \textbf{Recurrent Neural Networks (RNN)}: The RNN model consists of one bi-directional LSTM layer. The bi-directional LSTM train two LSTMs on the input sequence. The first one examines the input sequence in forward order, and the second one examines the input sequence backward and then combines the information from both ends to derive a single representation. This helps in learning a  better feature representation and capturing more sequential patterns from both directions.  The bi-directional LSTM layer is followed by a dropout layer and a fully connected layer.

 \item \textbf{Convolutional Recurrent Neural Networks (CRNN)}: For the final model, we used a combination of one-dimensional convolution layer and five bi-directional LSTM layers to create CRNNs (Convolutional Recurrent Neural Networks). The model uses a multi-scale convolutional layer with a kernel of 4 and 5  to extract multiple map features from the input text. Each kernel has a fixed length of 100.  We apply a nonlinear transformation using a Rectified Linear Unit (ReLU) \cite{nair2010rectified} to each feature map. Then, the max-pooling layer pools them separately to extract essential text features. Then the extracted features are concatenated and fed as input to  bi-LSTM layers. The bi-LSTM extracted the text features, and the output of bi-LSTM is fed to a fully connected layer.
 
\end{itemize}
Deep learning models have some advantages. For example, CNN automatically selects relevant words in tweets while the RNN-BiLSTM network captures the word patterns in tweets in two directions from right to left and vice versa and, unlike CNN, can manage the different lengths of tweets. The CRNN model combines the benefits of both networks.

\subsection{Evaluation Metric}
When dealing with imbalanced classification tasks, it is natural for the classifier to get biased toward the majority class. One of the most used techniques to solve the imbalance issue is to change the evaluation metric to a metric that tells a more truthful story. Therefore, when evaluating the models performance, we report different metrics including the Area Under the ROC Curve (AUC), precision, recall, and F1.  
The definition of these measurements is briefly outlined as follows:

 \textbf{The Area Under the ROC Curve (AUC)}: indicates the classifiers' ability to distinguish between classes through the probability curve ( ROC ). The AUC is defined as follows:
\begin{equation}
AUC = \sum_{i \in (TP+FP+FN+TN)}\frac{(TPR_i+TPR_{i-1}).(FPR_i+FPR_{i-1})}{2}
\end{equation}
 \textbf{Precision}: represent the percentage of positively classified tweets that actually correct. The precision is mathematically expressed as follows:
\begin{equation}
Precision= \frac{TP}{TP+FP}
\end{equation}
\textbf{Recall}: indicates the ability of the classifiers to classify all positive instances correctly. The recall is mathematically expressed  as follows:
\begin{equation}
Recall= \frac{TP}{TP+FN}
\end{equation}

Where $TP$ is the number of correctly identified tweets as misinformation,  $FP$ is the number of incorrectly identified tweets as misinformation, $TN$ is the number of correctly identified tweets as not misinformation, and $FN$ the number of incorrectly identified tweets as not misinformation.

 \textbf{F1 score} : indicates the weighted harmonic mean of both precision and recall. The F1 is mathematically expressed as follows:
\begin{equation}
 F1 score = 2\frac{Precision \cdot Recall}{Precision+Recall}
\end{equation}

\section{Experimental Results}
First, we shuffled the data to ensure that the model is not affected by order of the data. We randomly split the sample of 8786 annotated tweets into training and testing sets (80:20 splits) for the traditional classifiers and for the deep learning classifiers the sample was randomly split into training, testing, and validation sets (60:20:20 splits). 
\begin{table}[ht!]
\begin{center}
\begin{tabular}{ |c|c|c| }
\hline
\textbf{Classifiers} & \textbf{Hyper-parameters} \\ 
\hline
& \\
RF Classifier &\makecell{ Criterion:entropy, max\_depth:8, max\_features:log2,  n\_estimators:500,\\ class\_weight:balanced}\\
& \\
XGB Classifier & \makecell{Colsample\_bytree:0.8, gamma:2, max\_depth:5,  min\_child\_weight:1\\, subsample:1.0} \\
& \\
NB Classifier &  Alpha= 0.5, fit\_prior= True \\
& \\
SGD Classifie& \makecell{Alpha:0.0056, l1\_ratio:0.13, loss: modified\_huber, penalty:l2,\\max\_iter:6000 , class\_weight:balanced}\\
& \\
SVC Classifier & C:1, gamma:1, kernel:rbf, probability:True , class\_weight:balanced\\
& \\
\hline
\end{tabular}
\end{center}
\caption{Hyper-parameter Settings for Traditional Classifiers}
\label{tb1:Tsettings}
\end{table}


\begin{table}[ht!]
\begin{center}
\begin{adjustbox}{width=\textwidth}
\begin{tabular}{ |c|c|c| }
\hline
\textbf{Classifiers} & \textbf{Hyper-parameters} \\ 
\hline
& \\
\textbf{Cross Entropy Loss :  } & \\
& \\
CNN & \makecell{ max feature :3194 , max feature(\textsc{FastText}): 26275 , max feature(word2vec): 247180 , \\ dropout : 0.5  , kernel size: 4 \& 5 ,filter: 100, \\ epochs:500 , batch size: 32 , embedding size :200 }\\ 
& \\
RNN & \makecell{ max feature :3194 , max feature(\textsc{FastText} ):26275 , max feature(word2vec ):247180 ,\\ number of hidden node : 1 , dropout : 0.5  , epochs:500 ,\\ batch size: 32 , embedding size :200 } \\
& \\
CRNN & \makecell{ max feature :3194 ,  max feature(\textsc{FastText}):26275 , max feature(word2vec ):247180 , \\ number of hidden node : 5 , dropout : 0.5  ,kernel size: 4 \& 5 ,\\ filter: 100, epochs:500 ,\\  batch size: 32 , embedding size :200} \\

\textbf{AUCPR Loss  :  } & \\
 &  \\
CNN & \makecell{  max feature(\textsc{FastText} ): 26275 , max feature(word2vec ): 247180 , \\ dropout : 0.5  ,  kernel size: 4 \& 5 ,filter: 100,\\ epochs:600 , batch size: 32 , embedding size :200}\\ &\\

RNN & \makecell{ max feature(\textsc{FastText} ):26275 , max feature(word2vec ):247180 , \\ number of hidden node : 1 , dropout : 0.5  , epochs:600 ,\\ batch size: 32 , embedding size :200}\\
& \\
CRNN & \makecell{ max feature(\textsc{FastText} ):26275 , max feature(word2vec ):247180 , \\ number of hidden node : 5 , dropout : 0.5  ,kernel size: 4 \& 5  ,\\ filter: 100, epochs:600 ,\\ batch size: 32 , embedding size :200} \\
& \\
\hline
\end{tabular}
\end{adjustbox}
\end{center}
\caption{ Hyper-parameter Settings for Deep Learning Classifiers.}
\label{tb2:dlsettings}
\end{table}

\subsection{Analysis of the traditional classifiers }
Since our dataset is imbalanced, we constructed a grid search to find the best hyper-parameters and maximize the AUC score. In the grid-search function, we chose AUC as the scoring parameter with 5-fold cross-validation. The model trains in each fold with all training data using all parameter combinations. To find the optimum parameter for the fold, each trained model is evaluated on the validation set. Then the trained model with the optimum parameters is used on the test set. This procedure is repeated until the model that maximizes the AUC score is found. \Cref{tb1:Tsettings}  show the hyper-parameter settings for traditional classifiers.  


\begin{table}[ht!]
\begin{center}
\begin{adjustbox}{width=\textwidth}
\begin{tabular}{| *{11}{c|} }
    \hline
\multirow{3}{*}{Classifiers} & \multicolumn{10}{|c|}{Features} \\
\cline{2-11}
&  \multicolumn{5}{c|}{TF-IDF Word Level} &
   \multicolumn{5}{c|}{TF-IDF Ngram}            \\             
\cline{2-11}
 &  Accuracy  &   AUC Score  & Precision & Recall & F1 Score &
    Accuracy  &    AUC Score & Precision & Recall & F1 Score 
       \\
\hline
RF Classifier  &
80.7\%  &  80.3\%  & 0.37 & 0.57 & 0.45 &
78.2\%  &  75.8 \% & 0.32 & 0.54 & 0.41 \\
\hline
XGB Classifier &
87.5\% &  80.6\% & 0.73 & 0.16 & 0.26 & 
87.0\% & 70.9\%  & 0.73 & 0.10 & 0.18  \\
\hline
NB Classifier &  
87.9\%  &  80.9\% & 0.70 & 0.22 & 0.34  & 
86.7\%  &  78.1\% & 0.55 &  0.24 & 0.34  \\
\hline
SGD Classifier &  
79.9\%  &  83.3\%  & 0.37  & 0.68 & 0.48 &
79.4\%  &  78.9\%  & 0.35  & 0.54 & 0.44 \\
\hline
SVC Classifier &  
87.8\%  &  82.9\%  & 0.59 & 0.39 & 0.47 &  
82.0\%  &  75.8\%  & 0.38 & 0.50 & 0.43 \\
\hline
\end{tabular}
\end{adjustbox}
\end{center}
\caption{Traditional Classifier Overall Performance } 
\label{tb3:TFresult}
\end{table}
We trained the traditional classifiers using unigram and n-gram  TF-IDF   feature representations. The n-gram was a combination of bigrams and trigrams. We also report the results based on word2vec and \textsc{FastText} embeddings methods. \Cref{tb3:TFresult} shows the Accuracy, AUC, Precision, Recall, and F1 measure results of the traditional classifiers for unigram and n-gram TF-IDF feature representations. The SVM classifier with unigram feature representations achieved the best accuracy of 87.8\%, AUC score of 82.9\%, and F-measure of 0.47, while the SGD classifier reached the highest recall 0.68 and the best AUC score of 83.3\% with unigram feature representations. For the precision, the XGB classifier achieved the highest precision, 0.73 for both unigram and n-gram. 
The results indicate that N-gram's size influences the accuracy rate with different classifiers. All the classifiers achieved the highest performance when using TF-IDF unigram.
\begin{table}[ht!]
\begin{center}
\begin{adjustbox}{width=\textwidth}
\begin{tabular}{| *{11}{c|} }
    \hline
\multirow{3}{*}{Classifiers} & \multicolumn{10}{|c|}{Features} \\
\cline{2-11}
& 
   \multicolumn{5}{c|}{word2vec}          &
   \multicolumn{5}{c|}{\textsc{FastText}}          \\             
\cline{2-11}
 &  Accuracy  &   AUC Score  & Precision & Recall & F1 Score &
    Accuracy  &    AUC Score & Precision & Recall & F1 Score   \\
\hline
RF Classifier  &
83.3\%  & 83.6\%   & 0.47 & 0.56 & 0.51 & 
84.3\%  &  84.3\%  & 0.50 & 0.53 & 0.52  \\
\hline
XGB Classifier &
86.2\% & 85.4\% & 0.67 & 0.25 & 0.37  &
86.8\% & 85.4\% & 0.72 & 0.27 & 0.39 \\
\hline
NB Classifier &  
74.4\%       & 81.2\%        & 0.35     & 0.741      & 0.47  & 
73.4        & 80.4        & 0.33     &0.69      & 0.45  \\
\hline
SGD Classifier &  
74.0\%  &  81.0\%  & 0.34  & 0.71 & 0.46 &
73.8\%   & 81.4\%   & 0.34  & 0.74 & 0.47 \\
\hline
SVC Classifier &  
76.6\%  & 84.2\%  & 0.38 & 0.81 & 0.52 & 
77.8\%   & 85.3 \%  & 0.40 & 0.80 & 0.53 \\
\hline
\end{tabular}
\end{adjustbox}
\end{center}
\caption{Traditional Classifier Overall  Performance based on word embedding methods}
\label{tb5:emresult}
\end{table}

\begin{figure}
\centering
\begin{subfigure}{.50\linewidth}
\centering
\includegraphics[width=.9\linewidth]{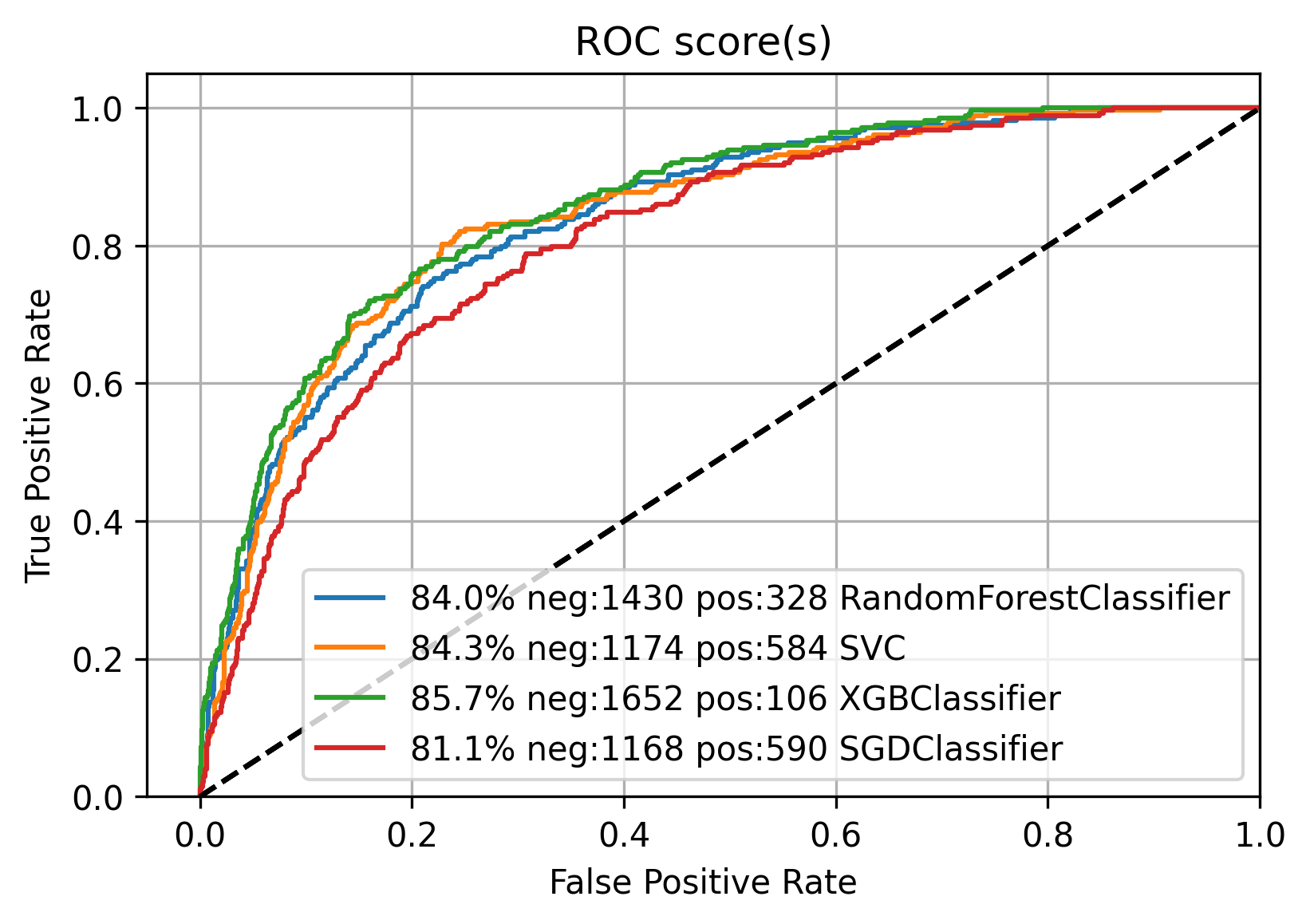}
\caption{}
\label{sfig:TC_word2vec_a}
\end{subfigure}%
\begin{subfigure}{.50\linewidth}
\centering
\includegraphics[width=.9\linewidth]{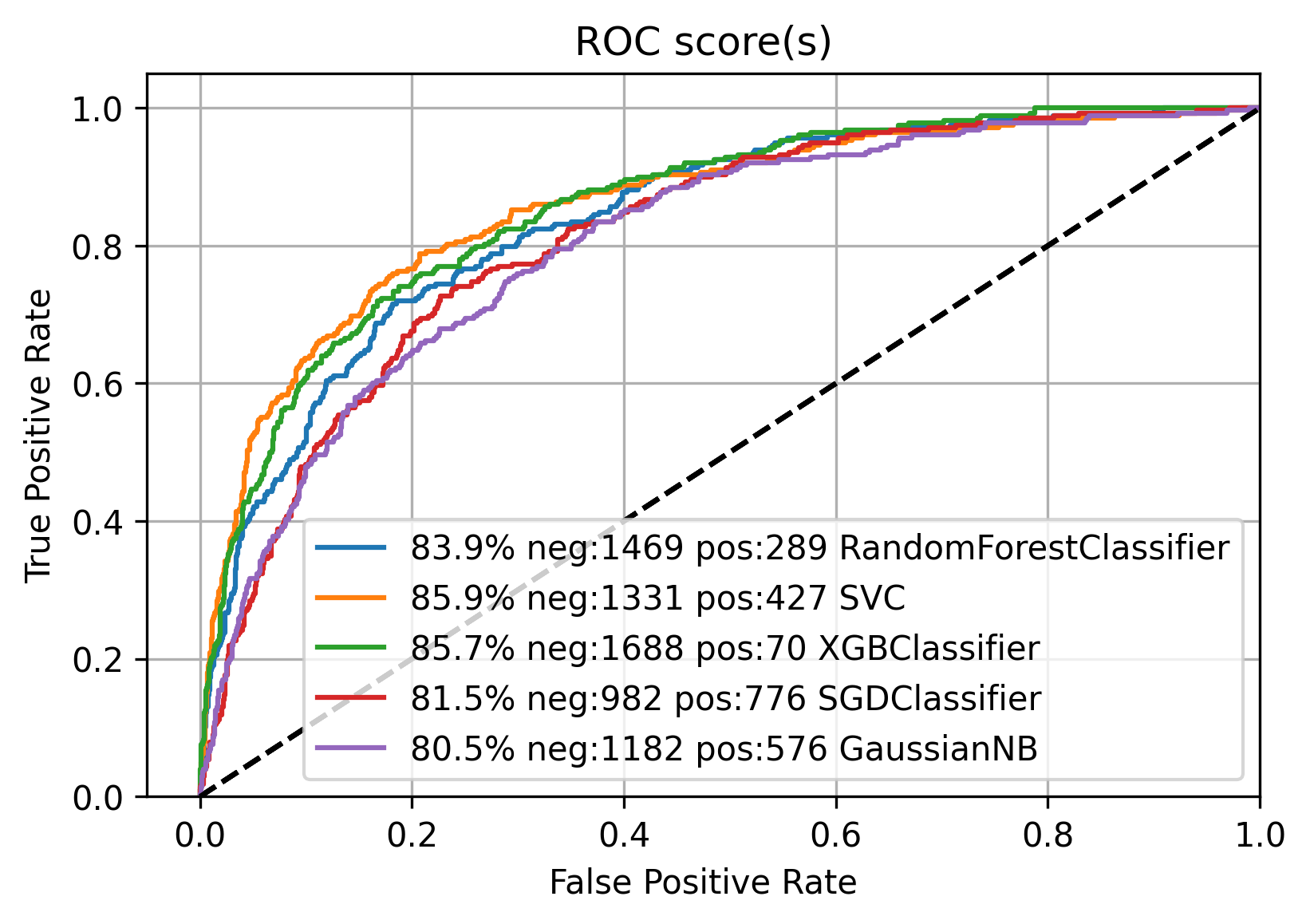}
\caption{}
\label{sfig:TC_FastText_b}
\end{subfigure}
\caption{Visualization of the ROC curves of traditional classifiers using (a) word2vec and (b) \textsc{FastText} word embedding techniques. }
\label{fig:tc_ROC}
\end{figure}

Using Both word2vec and \textsc{FastText} word embeddings results in slightly higher  AUC  and F1 scores. The overall AUC is increased by 1 to 5 points, as shown in \Cref{tb5:emresult}. Nevertheless, almost all classifier's performance improved with the trained word embeddings except for the SGD, where the AUC score decreased by 1 to 2 points. The highest AUC Score was generated by XGB classifier using both word embedding methods. However, the XGB classifier with \textsc{FastText} performs the best among the traditional classifiers, giving as much as 85.4\% AUC Score as well as second best precision of 0.72  an F1 score of 0.39, which signifies that the predication by XGB classifier is much better than all other classifiers. Followed by the SVC  classifier with a close AUC Score of  85.3 \%  and 0.80 recall with highest F1 score  0.53 among all classifiers. The ROC curve generated by the traditional classifiers using both word embedding methods are shown in \Cref{fig:tc_ROC}. 

\subsection{Analysis of the deep learning classifiers}
We trained the deep learning classifiers using adam optimizer to learn model parameters with  varying learning rates, a batch size of 32, and 500 epochs to optimize the cross-entropy loss. \Cref{tb2:dlsettings} shows the hyper-parameter settings for the deep learning classifiers. We reported the results with and without the pre-trained word embeddings. Without the pre-trained word embeddings, the accuracy of the CNN, RNN, and CRNN, 85.0\%, 84.3\%, 85.3\%, respectfully, and the AUC score was 50\%, 50\%, 50\%, as shown in \Cref{tb7:dlresult}. The deep learning classifiers generalize toward the majority class. With the pre-trained word2vec and \textsc{FastText} embeddings, the classifiers performance increased. The CRNN  has the best improvement  with   word2vec embedding as it outperformed all other classifiers with the highest AUC, precision, recall, and F1 score. Whereas for the CNN and RNN have the best improvement with the \textsc{FastText} embedding, while the CRNN showed the worse performance with the \textsc{FastText} embedding. 
The most significant improvement was with \textsc{FastText} embedding and the CNN  classifier's AUC score, which improved about  37.8\% compared to the previous results.

\begin{table}[ht]
\begin{center}
\begin{adjustbox}{width=\textwidth}
\begin{tabular}{|c|c|c|c|c|c|c |}

 \cline{1-7}
  \textbf{embeding method} &
   \textbf{Classifier} &
   \textbf{Accuracy}     & 
   \textbf{AUC Score} &
   \textbf{Precision} &
   \textbf{Recall} &
   \textbf{F1 Score}  \\  
\cline{1-7}
 \multirow{3}{*}{Without pre-trained word embedding}  & CNN & 85.0\% & 50 \% & 0 & 0 & 0 \\
  & RNN &  84.3\% &  50 \% & 0 & 0 & 0\\ 
  & CRNN & 85.3\% & 50 \% & 0 & 0 & 0\\
 \cline{1-7}
  \multirow{3}{*}{\textsc{FastText}} & CNN & 70\% & 68.6 \% & 0.75 & 0.43 & 0.54 \\
  & RNN &  83.6\% &  64.3 \% & 1 & 0.28 & 0.44\\ 
  & CRNN & 85.0\% & 50 \% & 0 & 0 & 0\\
  \cline{1-7}
    \multirow{3}{*}{word2vec} & CNN & 85.7\% & 57.1 \% & 1 & 0.14 & 0.25 \\
  & RNN &  82.9\% &  57.1 \% & 1 & 0.14 & 0.25\\ 
  & CRNN & 85.3\% & 64.3 \% &1 & 0.29 & 0.44\\
  \cline{1-7}
\end{tabular}
\end{adjustbox}
\end{center}
\caption{Deep learning Classifier Overall  Performance }
\label{tb7:dlresult}
\end{table}
To handle the imbalanced data set and further improve the classifier performance, we conducted a second experiment. We trained the classifiers using AUCPRLoss loss function that optimize for AUC based on \cite{eban2017scalable}, they introduced simple building block bounds that provide a unified framework for efficient, salable optimization of a wide range of objectives, including directly optimizing AUC. We used Adam's optimizer with varying learning rates,  a batch size of 32, and 600 epochs. 
\Cref{tb7:dlresult} shows the deep learning classifiers results After AUC optimization.
Without the pre-trained word embeddings, the performance of the classifier remarkably improved for all classifiers. The loss function had improved the AUC score about  26.8\% for all classifiers, and the models were able to detect the minority class. 

\begin{table}[ht]
\begin{center}
\begin{adjustbox}{width=\textwidth}
\begin{tabular}{|c|c|c|c|c|c|c| }

 \cline{1-7}
  \textbf{embeding method} &
   \textbf{Classifier} &
   \textbf{Accuracy}     & 
   \textbf{AUC Score} &
   \textbf{Precision} &
   \textbf{Recall} &
   \textbf{F1 Score}  \\  
\cline{1-7}
 \multirow{3}{*}{Without pre-trained word embedding} & CNN & 84.2\% & 64.3 \% & 1 & 0.29 & 0.44 \\
  & RNN &  81.6\% &  64.3 \%& 1 & 0.29 & 0.44 \\
  & CRNN & 86.0\% &  64.3 \%& 1 & 0.29 & 0.44 \\
 \cline{1-7}
  \multirow{3}{*}{\textsc{FastText}} & CNN & 74.9\% & 54.3 \% & 0.50 & 0.14 & 0.22 \\
  & RNN &  83.5\% &  64.3 \% & 1 & 0.14 & 0.25\\ 
  & CRNN & 85.0\% & 50 \% & 0 & 0 & 0\\
  \cline{1-7}
    \multirow{3}{*}{word2vec} & CNN & 68.4\% & 61.8\% & 0.67 & 0.29 & 0.40\\
  & RNN &  84.5\% & 57.1 \% & 1 & 0.14 & 0.25\\ 
  & CRNN & 85.1\% & 64.3 \% & 1 & 0.29 & 0.44\\
  \cline{1-7}
\end{tabular}
\end{adjustbox}
\end{center}
\caption{Deep learning Classifier Overall  Performance  after AUC Score Optimization}
\label{tb8:AucDlResult}
\end{table}

Using the pre-trained word embeddings with AUCPRLoss  function seems  to improve only some of classifiers. Using word2vec improve the CNN classifer by  4.7 points. However, further hyper tuning the classifiers parameter may increase the performance. \Cref{tb8:AucDlResult} shows the overall performance for deep learning classifiers after optimizing the AUC.  Among the deep learning classifiers, The CNN with \textsc{FastText} embeddings with cross-entropy loss achieved the best performance overall with the highest  AUC score, precision, recall, and F1 score. 
\begin{figure}
\centering
\begin{subfigure}{.50\linewidth}
\centering
\includegraphics[width=.9\linewidth]{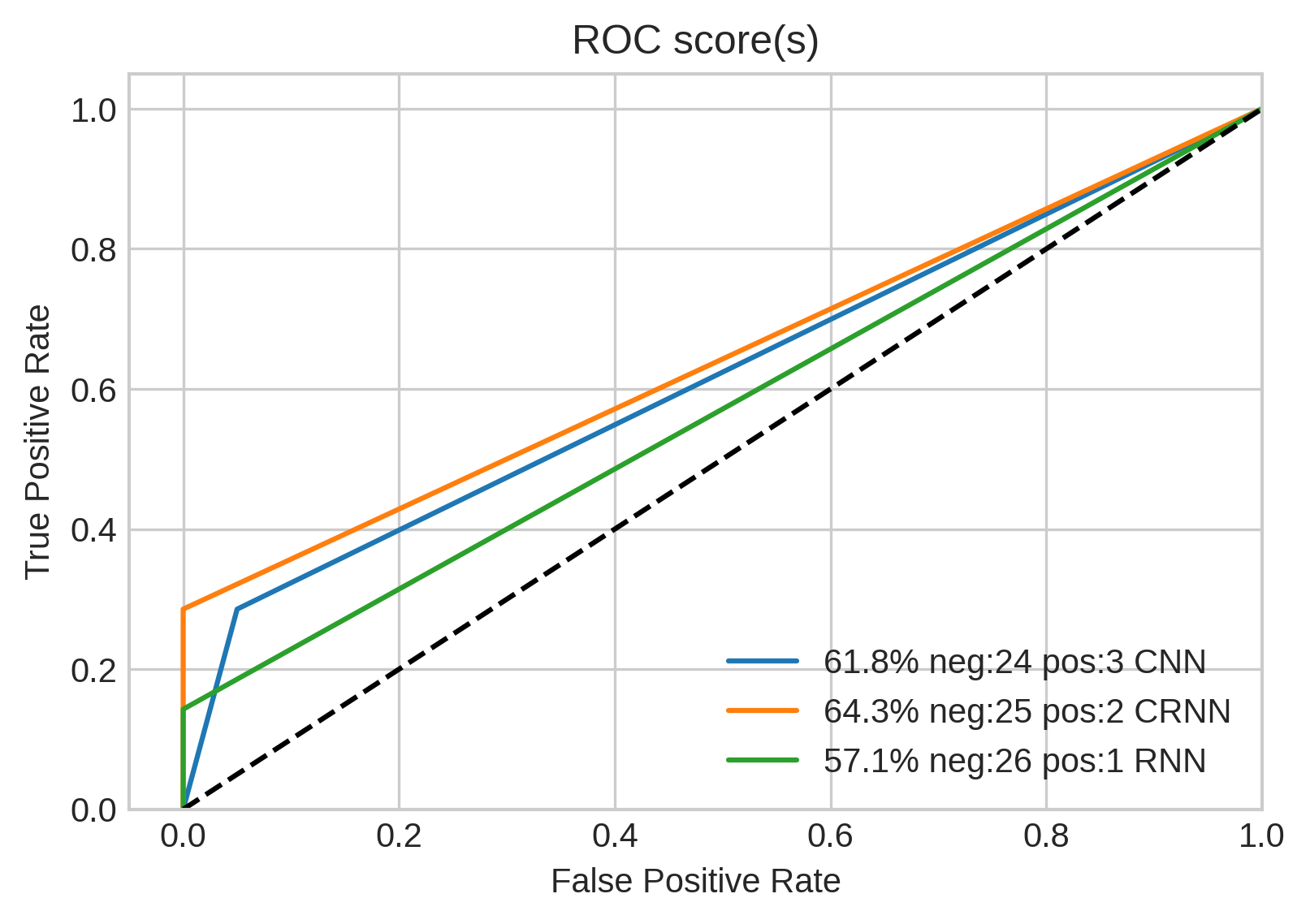}
\caption{}
\label{sfig:DL_word2vec_a}
\end{subfigure}%
\begin{subfigure}{.50\linewidth}
\centering
\includegraphics[width=.9\linewidth]{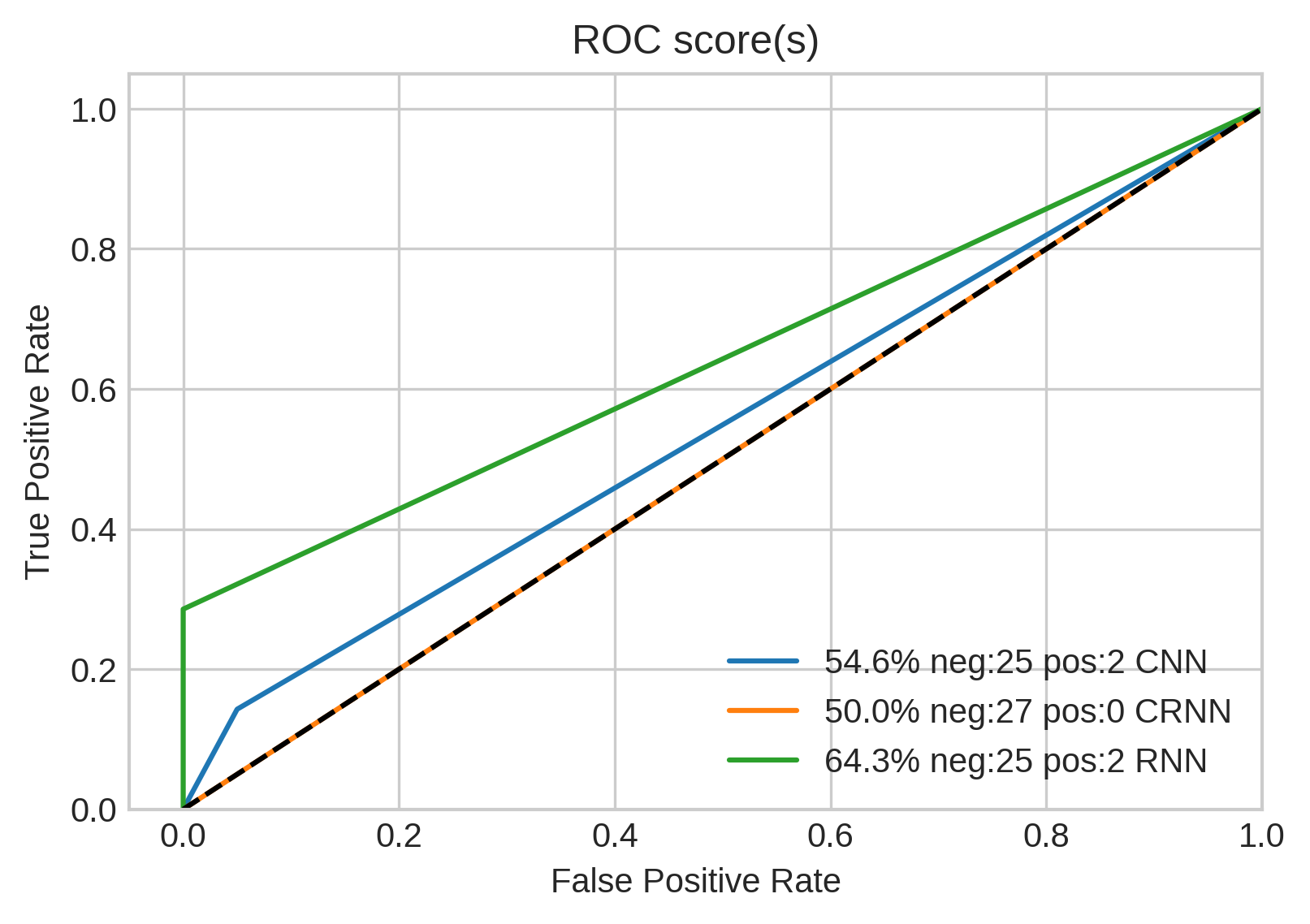}
\caption{}
\label{sfig:DL_FastText_b}
\end{subfigure}
\label{fig:DL_ROC}
\caption{ Visualization of the ROC curves of deep learning after AUC Score classifiers  using (a) word2vec and (b) \textsc{FastText} word embedding techniques.}
\end{figure}
\section{Discussion}
As a language, Arabic is a rich and complex language that has a vast vocabulary. It is also a highly morphological and derivative language. The complexity increases due to the informal nature of social media texts. There are two main forms of the  Arabic language present on social media: Modern Standard Arabic (MSA) and Dialectical Arabic (DA). Where the MSA is used for formal writing, and DA is used for informal daily communication. Nevertheless, the latter is the most common form.

Further complicating matters is that the Arabic language has many different dialects that people use in social media. The different dialects are one of the reasons for introducing many new words into any language, especially stop words \cite{alharbi2019identifying}. Another challenge is the diacritics used in the Arabic orthographic system. Diacritics are used to represent small vowels and to clarify the meanings of words. There are thirteen diacritics in the Arabic language \cite{almuzaini2020impact}. Many Arabic vocabularies has more than one meaning based on diacritics. 
Such as  \<حسب> , which can mean thought or counting, depending on the context. Furthermore, there are Arabic vocabularies that have different meaning based on the context, like the word \<عام>, which can mean public or year based on the context \cite{almuzaini2020impact}. In most of the Arabic tweets, the text is written without diacritics, where the reader is supposed to understand the purpose of the meaning. However, this does not apply to machines. In addition, the Arabic language contains a lot of grammatical rules that change the shape and meaning of the words.

Despite all these challenges, the classifiers showed promising results in distinguishing COVID-19 misinformation in Arabic tweets, which means that available machine learning methods can deliver high-performance and promising classifiers using an imbalanced dataset of tweets. Compared to deep learning, traditional classifiers have better performance with higher AUC values. Based on the experimental results, it is evident that feature selection can be an effective technique to improve traditional classifiers' performance.

Although the deep learning models are biased toward the minority class, the models' performance can increase using pre-train word embedding or by optimizing the AUC score. Using pre-trained word embedding on the disease-specific dataset can be more accurate than using other generic pre-trained word embedding in detecting health misinformation \cite{khatua2019tale}. As shown by the results, all classifiers perform better with the help of pre-train embedding than without it.  The \textsc{FastText} word embedding improved all classifiers' performance except for CRNN and NB, While word2vec improve the results for CRNN.  
{The key difference between word2vec and \textsc{FastText} is that during the learning phase, \textsc{FastText} tackles each word as composed of character n-grams whereas word2vec tackles words as the smallest unit.  Arabic is a morphological rich language, in addition, the social media posts such as tweets usually include informal writing and could be written with a misspelling which creates ambiguity. For example, the word corona :\<كورونا> could be written with several spellings
(\<كارونا>, \< كورنا>,\< الكرونا>, \<كوورنا>,\<كروونا>).
The \textsc{FastText} trained the embeddings vectors on the subwords units which consider the language morphology \cite{bojanowski2017enriching}. Therefore, \textsc{FastText} deals with mispelling and allows learning meaningful representation for rare words while word2vec ignores them.  For this reason, we believe the \textsc{FastText} tweet vectors representation is better compared to the word2vec representation.  }
The  Area under the ROC Curve (AUC) measures how well a machine learning model distinguishes between positive and negative tweets. A few studies have shown that optimizing the AUC is extremely useful for evaluating the classifier when class distributions are heavily imbalanced
\cite{bradley1997use,sulam2017maximizing}.  Our results confirm that maximizing the AUC score improves some of classifiers' results for imbalanced datasets.  It increases the ability of the classifiers to recognizes the minority classes. 
Nevertheless, optimizing for AUC is difficult because it requires dataset sorting, which makes it relatively expensive. As well, AUC is not continuous in the training set and, hence, most studies optimize for a variant of AUC that is differentiable. Many methods have been developed that directly optimize the AUC during the training of the classifiers \cite{wang2020efficient,liu2019stochastic}. 
Future studies are needed to investigate the impact of different AUC optimization techniques for detecting the misinformation.

{While the proposed dataset covers a diverse range of misinformation content, one limitation is that our work have been limited to tweets disseminated during March and April 2020. This is largely motivated by the fact that most of the Arabic-speaking countries reported their first confirmed cases of COVID-19 during March, 2020. Due to the lack of proper awareness and knowledge among people, the false information mostly spread at the early stage of the pandemic. Experimenting with a larger datasets spanning a longer duration (e.g., 5 months) would be useful to extend our work and validate it. }

\begin{figure}
\centering
\begin{subfigure}{.50\linewidth}
\centering
\includegraphics[width=.9\linewidth]{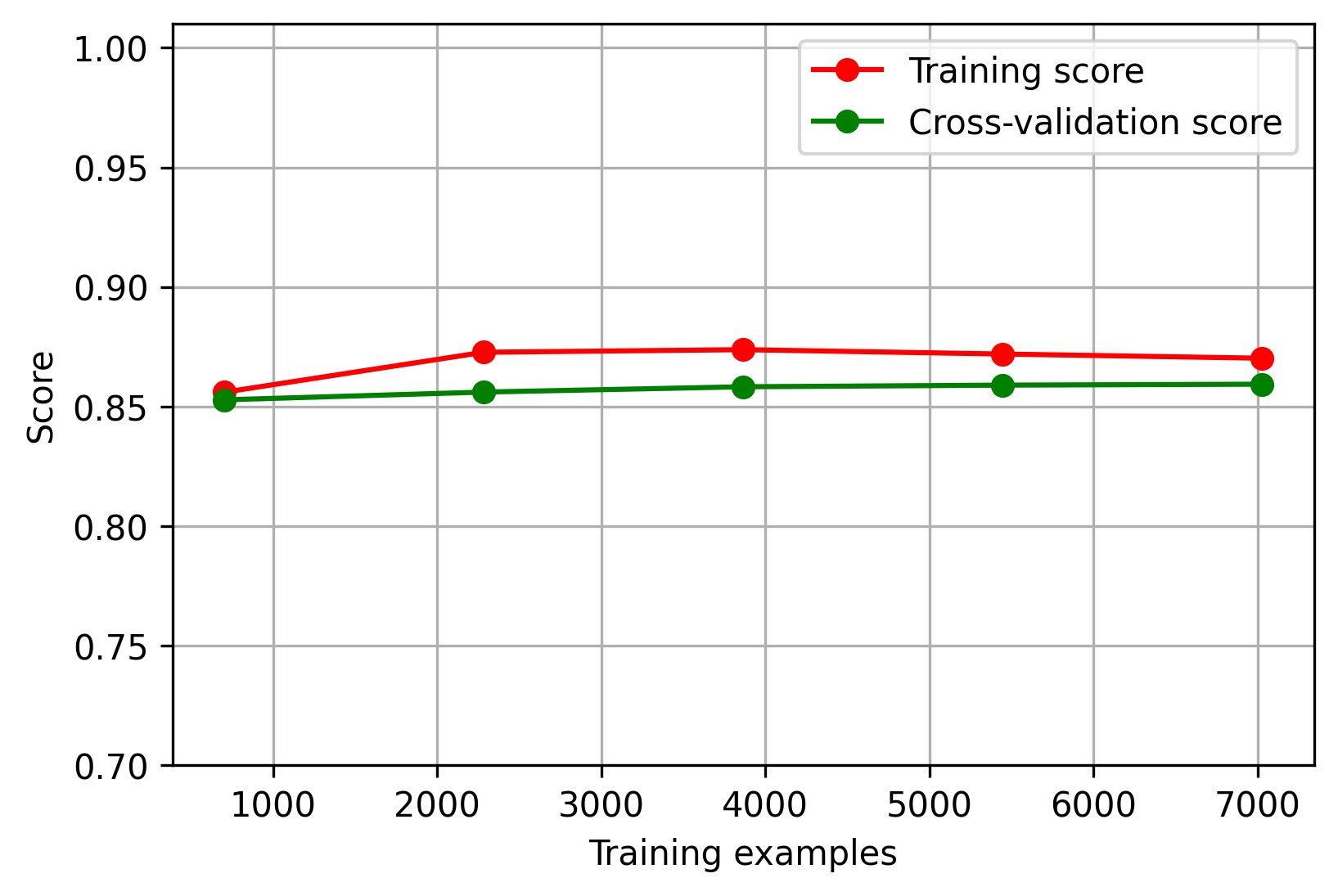}
\caption{learning curves for XGB classifier}
\label{sfig:lc_xgb}
\end{subfigure}%
\begin{subfigure}{.50\linewidth}
\centering
\includegraphics[width=.9\linewidth]{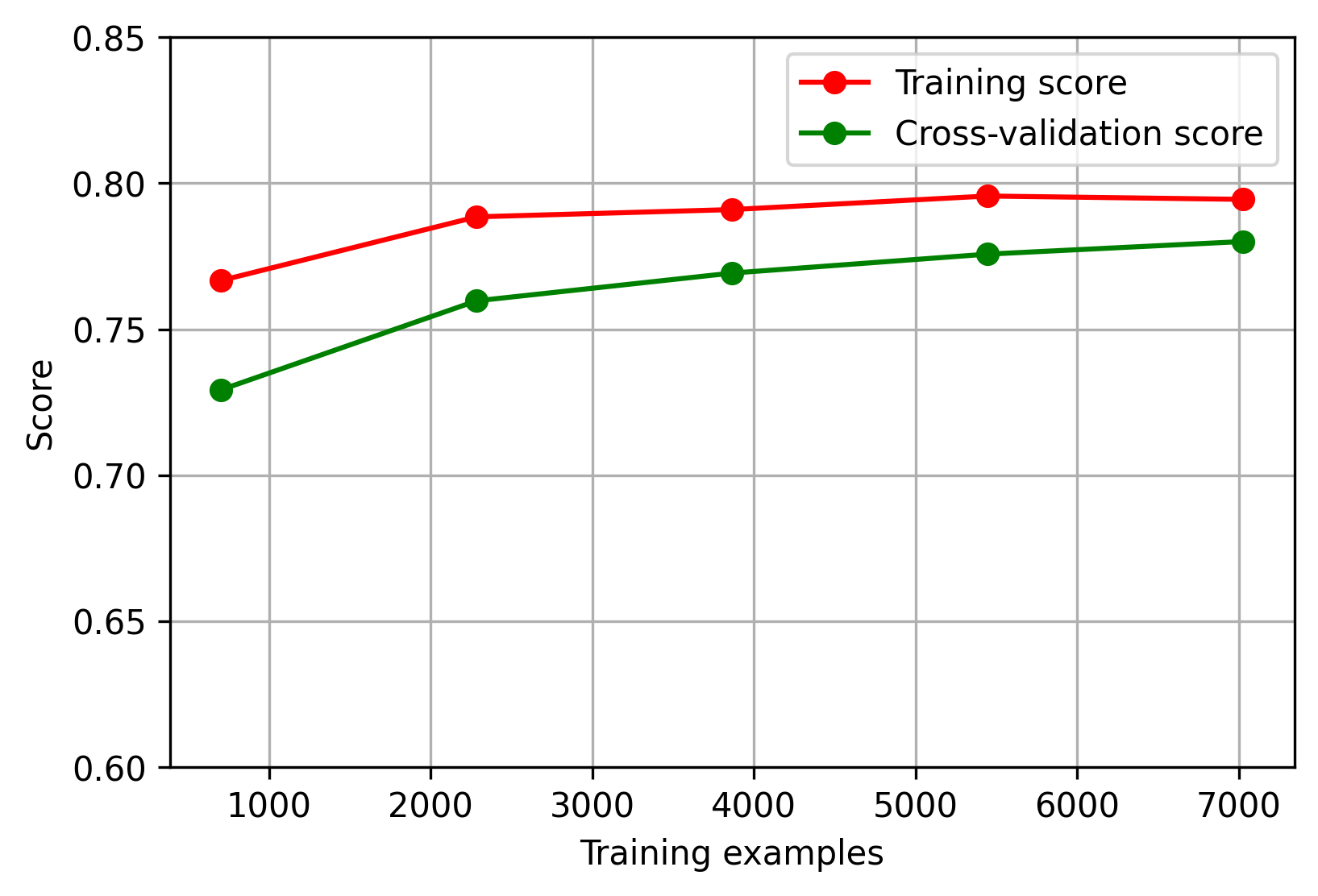}
\caption{learning curves for SVC classifier}
\label{sfig:lc_svc}
\end{subfigure}
\caption{learning curves for the classifiers}
\end{figure}

\section{Conclusion}
With the increase use of social media as a primary source of information, the distinction between correct and misleading information becomes very difficult and critical, especially during the ongoing COVID-19 pandemic. Many intervention strategies for COVID-19 depend on the quality and reliability of information shared between people. Several features in social media facilitate the spread of inaccurate information among users worldwide. Identifying and combating misinformation is therefore a critical task during pandemics. 

In this work, we conducted an extensive experiment using real misinformation content from Twitter. We examined different machine learning classifiers to identify Arabic misinformation related to COVID-19 automatically using an annotated dataset of 8786 tweets and employed word2vec and \textsc{FastText}.

 Our results show that using word embeddings will indeed enhance the performance of the classifier. \textsc{FastText} produces better results with traditional classifiers and the CNN, while word2vec allows for better results with the deep learning classifiers. Optimizing the AUC score improved the classifiers' performance and the ability to handle imbalanced datasets. The XGB classifier were shown to be capable of accurately identifying Arabic misinformation based solely on a tweet's text and it outperforms all other classifiers in terms of AUC, precision, recall, and F1.
 In the foreseeable future, we plan to improve deep learning classifiers by stacking multiple layers, and further optimize the hyperparameters, and possibly extend the study to include the very recent adabelief optimizer \cite{zhuang2020adabelief}. Finally, we indeed plan to consider other social networks that will help enriching our dataset and widening its applications. 
 
 \subsection*{Acknowledgement}
This work was supported by King Abdulaziz City for Science and Technology. Grant Number: 5-20-01-007-0033.
\bibliographystyle{ieeetr} 
\bibliography{ref}

\begin{thebibliography}{10}

\bibitem{ghenai2017catching}
A.~Ghenai and Y.~Mejova, ``Catching zika fever: Application of crowdsourcing
  and machine learning for tracking health misinformation on twitter,'' {\em
  arXiv preprint arXiv:1707.03778}, 2017.

\bibitem{oyeyemi2014ebola}
S.~O. Oyeyemi, E.~Gabarron, and R.~Wynn, ``Ebola, twitter, and misinformation:
  a dangerous combination?,'' {\em Bmj}, vol.~349, p.~g6178, 2014.

\bibitem{infodemic}
{World Health Organization}, ``Managing the covid-19 infodemic: Promoting
  healthy behaviours and mitigating the harm from misinformation and
  disinformation.''
  https://www.who.int/news/item/23-09-2020-managing-the-covid-19-infodemic-promoting-healthy-behaviours-and-mitigating-the-harm-from-misinformation-and-disinformation,
  2020.
\newblock (accessed: 2020-10-12).

\bibitem{del2016spreading}
M.~Del~Vicario, A.~Bessi, F.~Zollo, F.~Petroni, A.~Scala, G.~Caldarelli, H.~E.
  Stanley, and W.~Quattrociocchi, ``The spreading of misinformation online,''
  {\em Proceedings of the National Academy of Sciences}, vol.~113, no.~3,
  pp.~554--559, 2016.

\bibitem{MIT}
{Peter Dizikes}, ``study: on twitter, false news travels faster than true
  stories.''
  https://news.mit.edu/2018/study-twitter-false-news-travels-faster-true-stories-0308,
  2020.
\newblock (accessed: 2020-10-12).

\bibitem{donovan2020here}
J.~Donovan, ``Here’s how social media can combat the coronavirus
  ‘infodemic’,'' 2020.

\bibitem{persily2020social}
N.~Persily and J.~A. Tucker, {\em Social Media and Democracy: The State of the
  Field, Prospects for Reform}.
\newblock Cambridge University Press, 2020.

\bibitem{wu2019misinformation}
L.~Wu, F.~Morstatter, K.~M. Carley, and H.~Liu, ``Misinformation in social
  media: definition, manipulation, and detection,'' {\em ACM SIGKDD
  Explorations Newsletter}, vol.~21, no.~2, pp.~80--90, 2019.

\bibitem{islam2020deep}
M.~R. Islam, S.~Liu, X.~Wang, and G.~Xu, ``Deep learning for misinformation
  detection on online social networks: a survey and new perspectives,'' {\em
  Social Network Analysis and Mining}, vol.~10, no.~1, pp.~1--20, 2020.

\bibitem{lewandowsky2012misinformation}
S.~Lewandowsky, U.~K. Ecker, C.~M. Seifert, N.~Schwarz, and J.~Cook,
  ``Misinformation and its correction: Continued influence and successful
  debiasing,'' {\em Psychological science in the public interest}, vol.~13,
  no.~3, pp.~106--131, 2012.

\bibitem{akhtar2018no}
M.~S. Akhtar, A.~Ekbal, S.~Narayan, V.~Singh, and E.~Cambria, ``No, that never
  happened!! investigating rumors on twitter,'' {\em IEEE Intelligent Systems},
  vol.~33, no.~5, pp.~8--15, 2018.

\bibitem{buntain2017automatically}
C.~Buntain and J.~Golbeck, ``Automatically identifying fake news in popular
  twitter threads,'' in {\em 2017 IEEE International Conference on Smart Cloud
  (SmartCloud)}, pp.~208--215, IEEE, 2017.

\bibitem{wang2010don}
A.~H. Wang, ``Don't follow me: Spam detection in twitter,'' in {\em 2010
  international conference on security and cryptography (SECRYPT)}, pp.~1--10,
  IEEE, 2010.

\bibitem{al2010measuring}
R.~M.~B. Al-Eidan, H.~S. Al-Khalifa, and A.~S. Al-Salman, ``Measuring the
  credibility of arabic text content in twitter,'' in {\em 2010 Fifth
  International Conference on Digital Information Management (ICDIM)},
  pp.~285--291, IEEE, 2010.

\bibitem{hassan2018supervised}
N.~Y. Hassan, W.~H. Gomaa, G.~A. Khoriba, and M.~H. Haggag, ``Supervised
  learning approach for twitter credibility detection,'' in {\em 2018 13th
  International Conference on Computer Engineering and Systems (ICCES)},
  pp.~196--201, IEEE, 2018.

\bibitem{jardaneh2019classifying}
G.~Jardaneh, H.~Abdelhaq, M.~Buzz, and D.~Johnson, ``Classifying arabic tweets
  based on credibility using content and user features,'' in {\em 2019 IEEE
  Jordan International Joint Conference on Electrical Engineering and
  Information Technology (JEEIT)}, pp.~596--601, IEEE, 2019.

\bibitem{el2017cat}
R.~El~Ballouli, W.~El-Hajj, A.~Ghandour, S.~Elbassuoni, H.~Hajj, and K.~Shaban,
  ``Cat: Credibility analysis of arabic content on twitter,'' in {\em
  Proceedings of the Third Arabic Natural Language Processing Workshop},
  pp.~62--71, 2017.

\bibitem{sabbeh2018arabic}
S.~F. SABBEH and S.~Y. BAATWAH, ``Arabic news credibility on twitter: An
  enhanced model using hybrid features.,'' {\em Journal of Theoretical \&
  Applied Information Technology}, vol.~96, no.~8, 2018.

\bibitem{mouty2019effect}
R.~Mouty and A.~Gazdar, ``The effect of the similarity between the two names of
  twitter users on the credibility of their publications,'' in {\em 2019 Joint
  8th International Conference on Informatics, Electronics \& Vision (ICIEV)
  and 2019 3rd International Conference on Imaging, Vision \& Pattern
  Recognition (icIVPR)}, pp.~196--201, IEEE, 2019.

\bibitem{hassan2020credibility}
N.~Hassan, W.~Gomaa, G.~Khoriba, and M.~Haggag, ``Credibility detection in
  twitter using word n-gram analysis and supervised machine learning
  techniques,'' {\em Int. J. Intell. Eng. Syst}, vol.~13, pp.~291--300, 2020.

\bibitem{alzanin2019rumor}
S.~M. Alzanin and A.~M. Azmi, ``Rumor detection in arabic tweets using
  semi-supervised and unsupervised expectation--maximization,'' {\em
  Knowledge-Based Systems}, vol.~185, p.~104945, 2019.

\bibitem{saeeddetecting}
F.~Saeed, M.~Al-Sarem, E.~A. Hezzam, and W.~M. Yafooz, ``Detecting
  health-related rumors on twitter using machine learning methods,''

\bibitem{leng2020analysis}
Y.~Leng, Y.~Zhai, S.~Sun, Y.~Wu, J.~Selzer, S.~Strover, J.~Fensel, A.~Pentland,
  and Y.~Ding, ``Analysis of misinformation during the covid-19 outbreak in
  china: cultural, social and political entanglements,'' {\em arXiv preprint
  arXiv:2005.10414}, 2020.

\bibitem{serrano2020nlp}
J.~C.~M. Serrano, O.~Papakyriakopoulos, and S.~Hegelich, ``Nlp-based feature
  extraction for the detection of covid-19 misinformation videos on youtube,''
  in {\em Proceedings of the 1st Workshop on NLP for COVID-19 at ACL 2020},
  2020.

\bibitem{ng2020pofma}
L.~H.~X. Ng and L.~J. Yuan, ``Is this pofma? analysing public opinion and
  misinformation in a covid-19 telegram group chat,'' {\em arXiv preprint
  arXiv:2010.10113}, 2020.

\bibitem{singh2020first}
L.~Singh, S.~Bansal, L.~Bode, C.~Budak, G.~Chi, K.~Kawintiranon, C.~Padden,
  R.~Vanarsdall, E.~Vraga, and Y.~Wang, ``A first look at covid-19 information
  and misinformation sharing on twitter,'' {\em arXiv preprint
  arXiv:2003.13907}, 2020.

\bibitem{mourad2020critical}
A.~Mourad, A.~Srour, H.~Harmanani, C.~Jenainatiy, and M.~Arafeh, ``Critical
  impact of social networks infodemic on defeating coronavirus covid-19
  pandemic: Twitter-based study and research directions,'' {\em arXiv preprint
  arXiv:2005.08820}, 2020.

\bibitem{medford2020infodemic}
R.~J. Medford, S.~N. Saleh, A.~Sumarsono, T.~M. Perl, and C.~U. Lehmann, ``An"
  infodemic": Leveraging high-volume twitter data to understand public
  sentiment for the covid-19 outbreak,'' {\em medRxiv}, 2020.

\bibitem{pulido2020covid}
C.~M. Pulido, B.~Villarejo-Carballido, G.~Redondo-Sama, and A.~G{\'o}mez,
  ``Covid-19 infodemic: More retweets for science-based information on
  coronavirus than for false information,'' {\em International Sociology},
  p.~0268580920914755, 2020.

\bibitem{shahi2020exploratory}
G.~K. Shahi, A.~Dirkson, and T.~A. Majchrzak, ``An exploratory study of
  covid-19 misinformation on twitter,'' {\em arXiv preprint arXiv:2005.05710},
  2020.

\bibitem{mcquillan2020cultural}
L.~McQuillan, E.~McAweeney, A.~Bargar, and A.~Ruch, ``Cultural convergence:
  Insights into the behavior of misinformation networks on twitter,'' {\em
  arXiv preprint arXiv:2007.03443}, 2020.

\bibitem{caldarelli2020analysis}
G.~Caldarelli, R.~De~Nicola, M.~Petrocchi, M.~Pratelli, and F.~Saracco,
  ``Analysis of online misinformation during the peak of the covid-19 pandemics
  in italy,'' {\em arXiv preprint arXiv:2010.01913}, 2020.

\bibitem{graham2020like}
T.~Graham, A.~Bruns, G.~Zhu, and R.~Campbell, ``Like a virus: The coordinated
  spread of coronavirus disinformation,'' 2020.

\bibitem{ferrara2020types}
E.~Ferrara, ``What types of covid-19 conspiracies are populated by twitter
  bots?,'' {\em First Monday}, 2020.

\bibitem{ding2020challenges}
K.~Ding, K.~Shu, Y.~Li, A.~Bhattacharjee, and H.~Liu, ``Challenges in combating
  covid-19 infodemic--data, tools, and ethics,'' {\em arXiv preprint
  arXiv:2005.13691}, 2020.

\bibitem{elhadad2020detecting}
M.~K. Elhadad, K.~F. Li, and F.~Gebali, ``Detecting misleading information on
  covid-19,'' {\em IEEE Access}, vol.~8, pp.~165201--165215, 2020.

\bibitem{al2020lies}
M.~S. Al-Rakhami and A.~M. Al-Amri, ``Lies kill, facts save: Detecting covid-19
  misinformation in twitter,'' {\em IEEE Access}, vol.~8, pp.~155961--155970,
  2020.

\bibitem{alam2020fighting}
F.~Alam, F.~Dalvi, S.~Shaar, N.~Durrani, H.~Mubarak, A.~Nikolov, G.~D.~S.
  Martino, A.~Abdelali, H.~Sajjad, K.~Darwish, {\em et~al.}, ``Fighting the
  covid-19 infodemic in social media: A holistic perspective and a call to
  arms,'' {\em arXiv preprint arXiv:2007.07996}, 2020.

\bibitem{alsudias2020covid}
L.~Alsudias and P.~Rayson, ``Covid-19 and arabic twitter: How can arab world
  governments and public health organizations learn from social media?,'' in
  {\em Proceedings of the 1st Workshop on NLP for COVID-19 at ACL 2020}, 2020.

\bibitem{mubarak2020arcorona}
H.~Mubarak and S.~Hassan, ``Arcorona: Analyzing arabic tweets in the early days
  of coronavirus (covid-19) pandemic,'' {\em arXiv preprint arXiv:2012.01462},
  2020.

\bibitem{TextBlob}
{}, ``Textblob: Simplified text processing.''
  https://textblob.readthedocs.io/en/dev/, 2020.
\newblock (accessed: 2020-10-12).

\bibitem{farasapy}
{}, ``farasapy.'' https://pypi.org/project/farasapy/, 2020.
\newblock (accessed: 2020-10-12).

\bibitem{alqurashi2020identifying}
S.~Alqurashi, A.~Alashaikh, and E.~Alanazi, ``Identifying information
  superspreaders of covid-19 from arabic tweets,'' {\em Preprints}, 2020.

\bibitem{salton1988term}
G.~Salton and C.~Buckley, ``Term-weighting approaches in automatic text
  retrieval,'' {\em Information processing \& management}, vol.~24, no.~5,
  pp.~513--523, 1988.

\bibitem{mikolov2013efficient}
T.~Mikolov, K.~Chen, G.~Corrado, and J.~Dean, ``Efficient estimation of word
  representations in vector space,'' {\em arXiv preprint arXiv:1301.3781},
  2013.

\bibitem{bojanowski2017enriching}
P.~Bojanowski, E.~Grave, A.~Joulin, and T.~Mikolov, ``Enriching word vectors
  with subword information,'' {\em Transactions of the Association for
  Computational Linguistics}, vol.~5, pp.~135--146, 2017.

\bibitem{elrazzaz2017methodical}
M.~Elrazzaz, S.~Elbassuoni, K.~Shaban, and C.~Helwe, ``Methodical evaluation of
  arabic word embeddings,'' in {\em Proceedings of the 55th Annual Meeting of
  the Association for Computational Linguistics (Volume 2: Short Papers)},
  pp.~454--458, 2017.

\bibitem{zahran2015word}
M.~A. Zahran, A.~Magooda, A.~Y. Mahgoub, H.~Raafat, M.~Rashwan, and A.~Atyia,
  ``Word representations in vector space and their applications for arabic,''
  in {\em International Conference on Intelligent Text Processing and
  Computational Linguistics}, pp.~430--443, Springer, 2015.

\bibitem{joulin2016FastText}
A.~Joulin, E.~Grave, P.~Bojanowski, M.~Douze, H.~J{\'e}gou, and T.~Mikolov,
  ``Fasttext. zip: Compressing text classification models,'' {\em arXiv
  preprint arXiv:1612.03651}, 2016.

\bibitem{joulin2016bag}
A.~Joulin, E.~Grave, P.~Bojanowski, and T.~Mikolov, ``Bag of tricks for
  efficient text classification,'' {\em arXiv preprint arXiv:1607.01759}, 2016.

\bibitem{agibetov2018fast}
A.~Agibetov, K.~Blagec, H.~Xu, and M.~Samwald, ``Fast and scalable neural
  embedding models for biomedical sentence classification,'' {\em BMC
  bioinformatics}, vol.~19, no.~1, p.~541, 2018.

\bibitem{rehurek_lrec}
R.~{\v R}eh{\r u}{\v r}ek and P.~Sojka, ``{Software Framework for Topic
  Modelling with Large Corpora},'' in {\em {Proceedings of the LREC 2010
  Workshop on New Challenges for NLP Frameworks}}, (Valletta, Malta),
  pp.~45--50, ELRA, May 2010.

\bibitem{yang2018using}
X.~Yang, C.~Macdonald, and I.~Ounis, ``Using word embeddings in twitter
  election classification,'' {\em Information Retrieval Journal}, vol.~21,
  no.~2-3, pp.~183--207, 2018.

\bibitem{varoquaux2015scikit}
G.~Varoquaux, L.~Buitinck, G.~Louppe, O.~Grisel, F.~Pedregosa, and A.~Mueller,
  ``Scikit-learn: Machine learning without learning the machinery,'' {\em
  GetMobile: Mobile Computing and Communications}, vol.~19, no.~1, pp.~29--33,
  2015.

\bibitem{paszke2019pytorch}
A.~Paszke, S.~Gross, F.~Massa, A.~Lerer, J.~Bradbury, G.~Chanan, T.~Killeen,
  Z.~Lin, N.~Gimelshein, L.~Antiga, {\em et~al.}, ``Pytorch: An imperative
  style, high-performance deep learning library,'' in {\em Advances in neural
  information processing systems}, pp.~8026--8037, 2019.

\bibitem{nair2010rectified}
V.~Nair and G.~E. Hinton, ``Rectified linear units improve restricted boltzmann
  machines,'' in {\em ICML}, 2010.

\bibitem{eban2017scalable}
E.~Eban, M.~Schain, A.~Mackey, A.~Gordon, R.~Rifkin, and G.~Elidan, ``Scalable
  learning of non-decomposable objectives,'' in {\em Artificial intelligence
  and statistics}, pp.~832--840, PMLR, 2017.

\bibitem{alharbi2019identifying}
F.~R. Alharbi and M.~B. Khan, ``Identifying comparative opinions in arabic text
  in social media using machine learning techniques,'' {\em SN Applied
  Sciences}, vol.~1, no.~3, p.~213, 2019.

\bibitem{almuzaini2020impact}
H.~A. Almuzaini and A.~M. Azmi, ``Impact of stemming and word embedding on deep
  learning-based arabic text categorization,'' {\em IEEE Access}, vol.~8,
  pp.~127913--127928, 2020.

\bibitem{khatua2019tale}
A.~Khatua, A.~Khatua, and E.~Cambria, ``A tale of two epidemics: Contextual
  word2vec for classifying twitter streams during outbreaks,'' {\em Information
  Processing \& Management}, vol.~56, no.~1, pp.~247--257, 2019.

\bibitem{bradley1997use}
A.~P. Bradley, ``The use of the area under the roc curve in the evaluation of
  machine learning algorithms,'' {\em Pattern recognition}, vol.~30, no.~7,
  pp.~1145--1159, 1997.

\bibitem{sulam2017maximizing}
J.~Sulam, R.~Ben-Ari, and P.~Kisilev, ``Maximizing auc with deep learning for
  classification of imbalanced mammogram datasets.,'' in {\em VCBM},
  pp.~131--135, 2017.

\bibitem{wang2020efficient}
Q.~Wang and A.~Guo, ``An efficient variance estimator of auc and its
  applications to binary classification,'' {\em Statistics in Medicine},
  vol.~39, no.~28, pp.~4281--4300, 2020.

\bibitem{liu2019stochastic}
M.~Liu, Z.~Yuan, Y.~Ying, and T.~Yang, ``Stochastic auc maximization with deep
  neural networks,'' {\em arXiv preprint arXiv:1908.10831}, 2019.

\bibitem{zhuang2020adabelief}
J.~Zhuang, T.~Tang, S.~Tatikonda, N.~Dvornek, Y.~Ding, X.~Papademetris, and
  J.~S. Duncan, ``Adabelief optimizer: Adapting stepsizes by the belief in
  observed gradients,'' {\em arXiv preprint arXiv:2010.07468}, 2020.

\end{thebibliography}
\end{document}